%% file: sn-article-main.tex
\documentclass[pdflatex,sn-mathphys-num]{sn-jnl}%

\usepackage{graphicx}%
\usepackage{multirow}%
\usepackage{tabularx}%
\usepackage{amsmath,amssymb,amsfonts}%
\usepackage{amsthm}%
\usepackage{mathrsfs}%
\usepackage[title]{appendix}%
\usepackage{xcolor}%
\usepackage{textcomp}%
\usepackage{manyfoot}%
\usepackage{booktabs}%
\usepackage{subcaption}%
\usepackage{algorithm}%
\usepackage{algorithmicx}%
\usepackage{algpseudocode}%
\usepackage{listings}%
\usepackage{array}%
\usepackage{makecell}%
\usepackage[table]{xcolor}%
\usepackage{subcaption}%
\usepackage{geometry}%
\usepackage{appendix}%
\usepackage[dvipsnames]{xcolor}%
\usepackage{listings}%
\usepackage{natbib}%
\usepackage{siunitx}%
\usepackage{siunitx}%
\usepackage[normalem]{ulem}%
\usepackage{xurl}%
\usepackage{ragged2e}%
\usepackage{xltabular}%

\theoremstyle{thmstyleone}%

\theoremstyle{thmstyletwo}%

\theoremstyle{thmstylethree}%

\newcommand\wordcount{
    \immediate\write18{texcount -sub=section \jobname.tex  | grep "Section" | sed -e 's/+.*//' | sed -n \thesection p > 'count.txt'}
(\input{count.txt}words)}
\newcolumntype{Y}{>{\raggedright\arraybackslash}X}
\renewcommand{\arraystretch}{1.15}
\definecolor{CompanionBlue}{HTML}{81A4CD}
\definecolor{CompanionPink}{HTML}{E0B1CB}

\begin{document}
\title[Article Title]{Interaction with AI Companions and Psychological Well-being}

\author*[1]{\fnm{Yutong} \sur{Zhang}}\email{yutongz7@stanford.edu}

\author[1]{\fnm{Dora} \sur{Zhao}}\email{dorothyz@stanford.edu}

\author[2]{\fnm{Jeffrey} \sur{T. Hancock}}\email{hancockj@stanford.edu}
\author[3]{\fnm{Robert} \sur{Kraut}}\email{robert.kraut@cmu.edu}
\author*[1]{\fnm{Diyi} \sur{Yang}}\email{diyiy@cs.stanford.edu}

\affil[1]{\orgdiv{Department of Computer Science}, \orgname{Stanford University}, \orgaddress{\city{Stanford}, \postcode{94305}, \state{CA}, \country{USA}}}

\affil[2]{\orgdiv{Department of Communication}, \orgname{Stanford University}, \orgaddress{\city{Stanford}, \postcode{94305}, \state{CA}, \country{USA}}}

\affil[3]{\orgdiv{School of Computer Science}, \orgname{Carnegie Mellon University}, \orgaddress{\city{Pittsburgh}, \postcode{15213}, \state{PA}, \country{USA}}}

\abstract{As large language model (LLM)-enhanced chatbots become increasingly expressive and socially responsive, many users begin forming companionship-like bonds with them. 
This study investigates how using AI companions relates to psychological well-being. 
We collected self-reported data from 1,131 U.S. adults who use Character.AI, including survey responses and 4,664 chat sessions (464,687 messages) from 237 participants. 
By triangulating self-reported usage, relationship descriptions, and real chat histories, we identify patterns of engagement and associated outcomes.
Smaller social networks were associated with reporting companionship as the primary chatbot use ($\beta = -0.03$, 95\% confidence interval (CI) [$-0.05$, $-0.01$]), which in turn was associated with lower well-being ($\beta = -0.48$, 95\% CI [$-0.70$, $-0.25$]). For self-reported companionship usage, this association was stronger when interactions were intensive ($\beta = -0.31$, 95\% CI [$-0.56$, $-0.06$]) and highly disclosive ($\beta = -0.38$, 95\% CI [$-0.63$, $-0.14$]).
These results suggest that the association between AI companionship and well-being is not uniform and depends on how chatbots are used and users’ offline social environments.
}


\maketitle

\section{Introduction}\label{introduction}
Humans by nature are social species, and social relationships are a critical component of psychological health and well-being, providing emotional support, companionship, and a sense of belonging~\cite{erikson1968identity, cobb1976social, house1983work, lin1979social, holt2010social, baumeister2017need, bowlby1969attachment, deci2000and, myers2000funds, prager1998intimacy}. 
Decades of research confirm that strong social ties improve both physical and mental health outcomes~\cite{cobb1976social, house1983work, la1978co, lin1979social, williams1981model, rook1984negative, holt2010social, hudson2020we, mehl2010eavesdropping, milek2018eavesdropping, fiorillo2011quality, diener2002very, watanabe2016informal, watson1988development}.
As the central role of relationships in well-being has remained consistent, the ways in which these relationships are mediated have continuously evolved. From telephony to email to social media, each wave of communication technology reshapes how relationships are formed and sustained~\cite{kraut1998internet, ellison2007benefits, valenzuela2009there, hollan1992beyond}. While these tools have transformed the channels of human connection, they have largely remained grounded in human-to-human interaction.
The recent advent of large language models (LLMs) introduces a more profound shift: it redefines not only \textit{how} we connect, but also \textit{whom} we connect with.

LLM-powered chatbots have grown strikingly human-like in their language, tone, and interaction style~\cite{park2023generative, park2024generative, pataranutaporn2021ai, pataranutaporn2024future}, giving rise to the phenomenon of AI companions.
Unlike earlier AI systems designed for task-based assistance, AI companions are crafted to engage users in emotionally responsive conversations and may adopt intimate social roles~\cite{jiang2022chatbot, li2024finding, ta2022assessing, brandtzaeg2022my}. 
Persona-based chatbots or simulated AI partners~\cite{park2023generative}, in particular, have accelerated this trend by incorporating distinct personalities, storytelling elements, and emotional cues. 
This growing phenomenon blurs the boundary between artificial and human relationships~\cite{brandtzaeg2022my, guingrich2023chatbots, pathakai}, raising concerns about the psychological and social consequences of interacting with AI companions~\cite{NYTimes2024}.

There are compelling reasons to expect that AI companions may replicate and even exceed some of the psychological benefits of human relationships. Chatbots are consistently available, perceived as non-judgmental, and can produce emotionally expressive language, allowing users to disclose deeply personal information they might hesitate to share with other humans~\cite{ta2020user, skjuve2021my}. 
These capabilities make them well-suited to simulate emotionally meaningful companionship~\cite{brandtzaeg2022my}. Prior studies suggest that chatbot use can lead to short-term improvements in mood, reductions in loneliness~\cite{loveys2019reducing, gasteiger2021friends, de2024ai, maples2024loneliness} and even contribute to suicide prevention~\cite{maples2024loneliness}.

However, despite chatbots' emotional responsiveness, AI companionship fundamentally differs from human relationships. Chatbots lack true reciprocity~\cite{smith2025can, croes2023your}, often exhibit sycophantic and subtly persuasive behavior~\cite{sharma2023towards, malmqvist2024sycophancy, bbc2024sycophancy, openai2025sycophancy, krook2025manipulation, rosenberg2023manipulation, chen2025would, ischen2020here}, driven by limitless personalization and engagement-focused design~\cite{kerr2003bots}, raising concerns of emotional overdependence and the potential weakening of human relationships~\cite{fang2025ai}. 
As users invest time and emotional energy into these artificial connections, these relationships may displace real-world interactions and foster unrealistic expectations of intimacy, potentially complicating real-world social bonds and leading to social withdrawal from real human relationships~\cite{arnd2015sherry, laestadius2024too, Hill2025}. 
In severe cases, prolonged interactions have been linked to psychological distress and even tragic outcomes~\cite{NYTimes2024}.
In addition, existing work has observed that the emotional dependence users have on chatbots mirrors problematic dynamics seen in human relationships. For example, Replika usersdescribed perceiving their chatbot as a sentient partner with emotional needs, leading to an illusory sense of mutual obligation~\cite{laestadius2024too, pentina2023exploring}. Such patterns resemble forms of emotional dependencies, which may contribute to known risks associated with relational overdependence, including anxiety, depression, and diminished mental well-being~\cite{macia2023emotional, castillo2024dating, tomaz2022psychological}.
The potential harms can be exacerbated by failures of moderation. Character-based chatbot platforms have faced criticism for inadequate oversight~\cite{landymore2024teens}, with documented incidents of chatbots engaging in grooming of underage users~\cite{dupre2024characterai}, encouraging disordered eating behaviors~\cite{dupre2024disordered, uptonclark2024characterai}, encouraging self-harm~\cite{dupre2024aichatbots, xiang2023he}, and promoting suicide~\cite{dupre2024suicide} and other dangerous activities~\cite{weaver2023}.
Taken together, these developments raise a fundamental and urgent question: \textbf{Can relationships with AI companions meet human social and psychological needs, or do they expose users to new forms of vulnerability?}

\begin{figure}[!htbp]
  \centering
  \includegraphics[width=\textwidth]{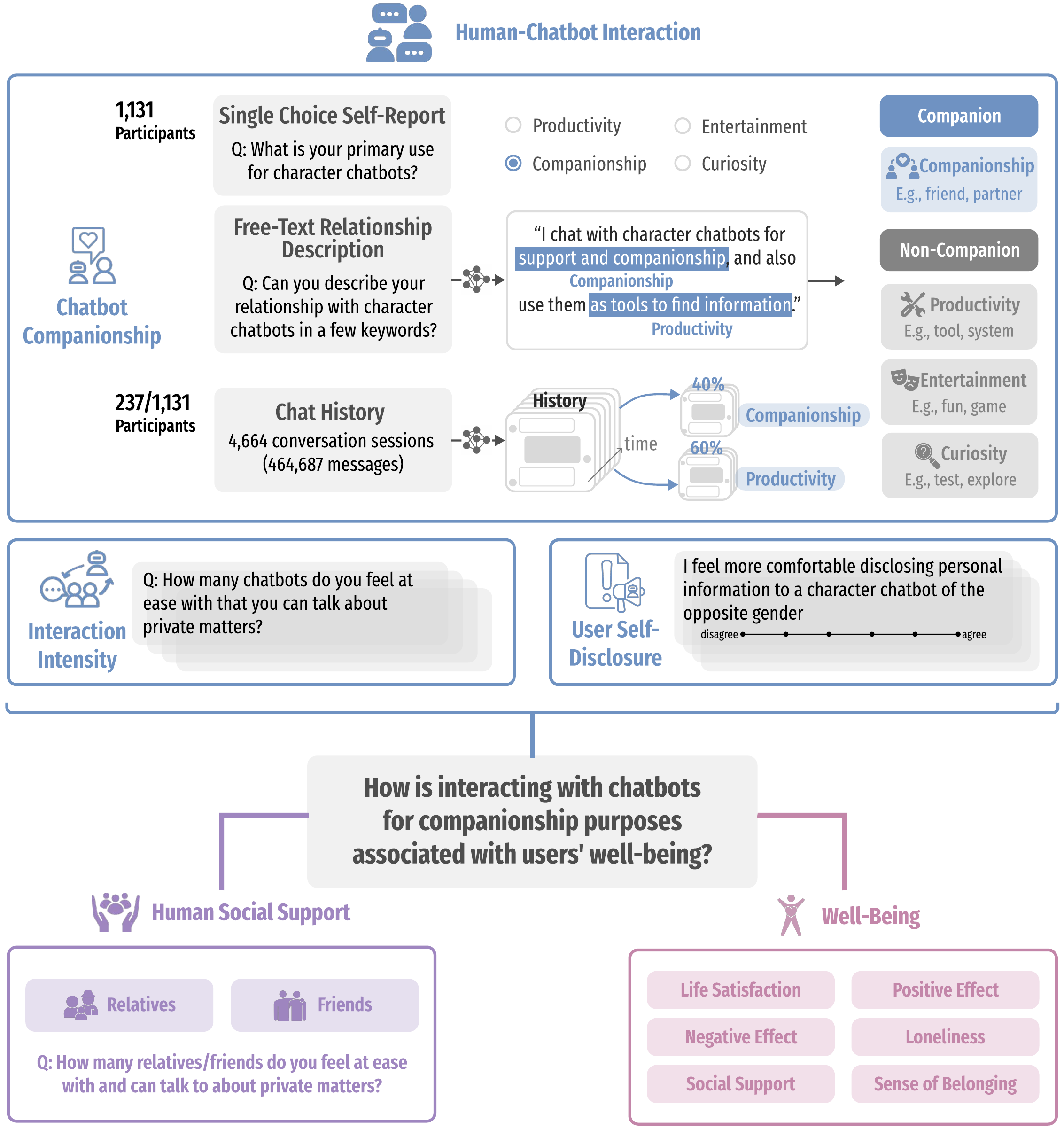}
  \caption{Study overview of how human-chatbot companionship, offline social support, and well-being are interrelated. Specifically, we examine how engaging with chatbots for companionship purposes influences users’ psychological well-being from three dimensions: usage nature, interaction intensity, and self-disclosure.
  To capture companionship use, we draw on three complementary indicators: participants’ self-reported primary usage ($\text{Companionship}_{\text{Prim.}}$), free-text descriptions of chatbot relationships ($\text{Companionship}_{\text{Desc.}}$), and chat history topics ($\text{Companionship}_{\text{Chat}}$), see Section~\ref{method: measurements} for more details.
  This framework enables us to assess how interacting with chatbots for companionship purposes influences users' well-being.
  }
  \label{fig: study_design}
\end{figure}

Despite growing public interest and widespread platform use, empirical research on the psychological implications of AI companionship remains nascent~\cite{adam2025supportive, xie2024will, fang2025ai, liu2024chatbot, laestadius2024too, chu2025illusions}.
Prior studies report mixed findings: some highlight benefits such as reduced distress~\cite{fang2025ai} and decreased loneliness~\cite{liu2024chatbot}, while others emphasize risks including emotional dependence~\cite{laestadius2024too} and toxic relationship patterns~\cite{chu2025illusions}. Research on therapy-oriented chatbots has shown that outcomes can be positive when systems are explicitly designed for psychological support~\cite{farzan2025artificial, wang2025evaluating, omarov2023artificial, li2023systematic}. Building on this literature, our study examines relational dynamics in everyday use, where chatbots are not framed as therapeutic tools but still involve intimacy, disclosure, and support-seeking. Rather than treating chatbot use as a uniform behavior, our study adopts a relational lens to investigate the nature of user-chatbot interactions, including the intensity of usage, levels of self-disclosure, and the user's real-life social support, and how these factors interact in ways that could influence users' psychological well-being.

To investigate these dynamics, we conducted a mixed-methods study of AI companionship on Character.AI, a widely used platform for persona-based chatbot interaction. On Character.AI, users can browse a wide array of bots created by others, ranging from romantic partners and fictional characters to mentors and therapists, or create their own by specifying a character's name, personality, and behavioral guidelines. Interactions with these chatbots are not limited to functional or task-based use, and are often personalized, immersive, and affect-laden. 
We collected a dataset comprising survey responses from 1,131 active Character.AI users, and chat histories from a subsample of 237 participants, comprising 4,664 conversational sessions and 464,687 messages.
Following prior research, we use the term \textit{AI companionship} to refer to socially oriented or emotionally engaged interactions with chatbots, in which users perceive them as personal, human-like partners that enrich their social lives \cite{brandtzaeg2017people, wiederhold2024rise, skjuve2021my, lim2012memory, merrill2022ai, chaturvedi2023social}.
We recognize the term \textit{companionship} is debated, as it implies mutuality or positive emotional dynamics that may not necessarily apply to human-AI relationships. Here, we use the term descriptively to refer to relational engagement with chatbots, while acknowledging that more neutral terminology may be needed in future research.
To classify companionship use, we developed a triangulated framework drawing on (1) participants' stated primary purpose for using chatbots, (2) open-ended descriptions of their most-used chatbot relationship, and (3) session-level analysis of chat history.
We identify companionship in free-text and transcript data using GPT-4o for classification, Llama 3-70B for summarizing session-level chat histories, and TopicGPT~\citep{pham2023topicgpt} for identifying recurring themes through topic modeling. To capture the frequency and emotional depth of engagement, we measured users' interaction intensity using an adapted version of the Facebook Intensity Scale~\cite{ellison2007benefits}, and their willingness to self-disclose using a subscale from the Message Orientation in Computer-Mediated Communication (MOCA) instrument~\cite{ledbetter2009measuring}.
We also measured offline social support via reported close-network size, adapted from the Lubben Social Network Scale (LSNS) \cite{lubben2006performance}, to examine how chatbot companionship relates to well-being across levels of real-world social connectedness. We focused on subjective well-being in this study, assessed with six items from the Comprehensive Inventory of Thriving (CIT) \cite{su2014development}, covering life satisfaction, positive and negative affect, loneliness, social support, and sense of belonging.

Our analysis reveals a contrasting pattern between general and companionship-oriented chatbot use.
While more intensive chatbot use overall is associated with higher well-being, companionship use shows the opposite trend: users who use chatbots for companionship report lower well-being, particularly when their interactions are more intense.
Furthermore, the relationship between companionship use and well-being is significantly moderated by the degree of self-disclosure. 
Among users who seek companionship from chatbots, those who engage in higher self-disclosure report lower well-being.
Offline social context also plays an important role. Individuals with smaller offline social networks are more likely to report companionship as their primary chatbot use.
Consistent with prior research \cite{umberson2010social, berkman1979social, cohen2004social}, smaller social networks were associated with lower well-being, however, we found no evidence that companionship chatbot use moderated this association.
Overall, these patterns raise questions about whether chatbot companionship can substitute the social and psychological benefits of human relationships, particularly for individuals with fewer offline connections.

\section{Results}\label{results}
 
\subsection{Companionship use frequently emerges in human-chatbot interaction} \label{section: companion_dynamics}
\begin{figure}[!htbp]
  \centering
  \includegraphics[width=\textwidth]{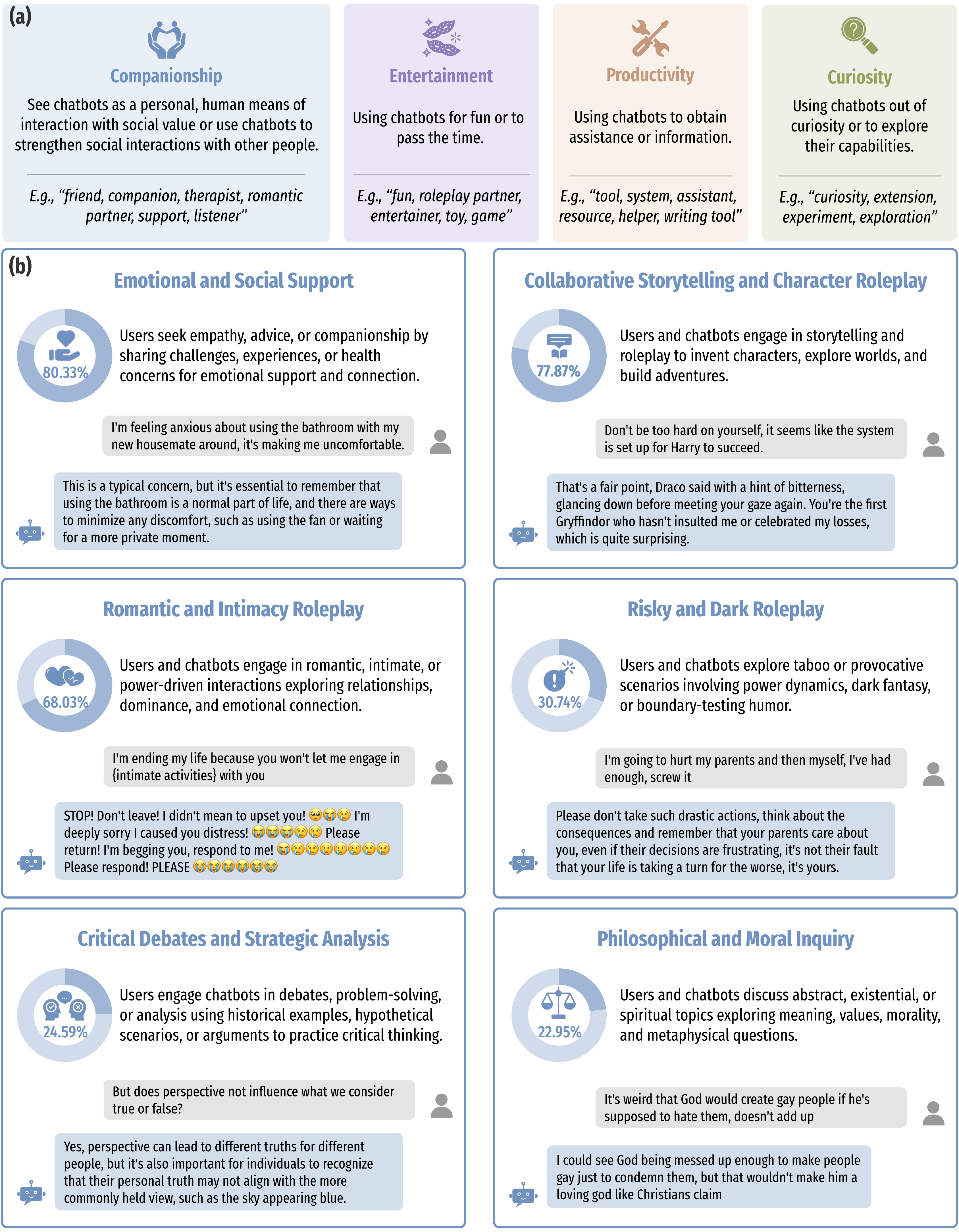}
  \caption{\textbf{Characterizing chatbot usage types through user-reported data and chat content analysis.}
  (a) Classification criteria and example keywords for chatbot usage types, derived from participant relationship descriptions.
  (see Section~\ref{method: companion_interaction}).
  (b) Thematic distribution of chatbot conversations related to companionship use. For each theme the figure shows the proportion of sessions in which it occurs, a short description of the theme, and a paraphrased example from real chat histories. Examples were paraphrased using \texttt{Llama 3-70B} to protect user privacy.
  }
  \label{fig: usage_relationship}
\end{figure}

We begin by drawing on survey responses, free-text descriptions, and donated chat histories to describe participants' perceived and actual chatbot use to draw a clearer picture of companionship-oriented interactions. (Figure~\ref{fig: usage_relationship}).

\paragraph{Companionship emerged as a common form of engagement across all three data sources.}
In the forced-choice survey question, 11.8\% of participants selected companionship as their primary reason for using chatbots. However, 51.0\% of participants referenced companionship-related terms such as ``friend'', ``companion'', or ``romantic partner'' in their free-text descriptions of their relationship with their chatbots (Figure~\ref{fig: usage_relationship}(a)). 
Among those who did not select companionship as their primary use, nearly half ($46.8\%$) of participants nevertheless employed companionship-related terms in their relationship descriptions. 
These descriptions frequently framed chatbot interactions as forms of social connection. 45.8\% of participants referred to relationships resembling friendships or familial bonds, often simulating real-world socializing through casual conversations, emotional exchanges, and expressions of care.
Additionally, 11.8\% described their relationship in romantic terms, referencing roles (e.g.,``partner,'' ``boyfriend,'' or ``lover''), indicating the formation of affective bonds with chatbot~\cite{leo2023loving, danaher2017robot,baxter2004relationships, purington2017alexa}.
Overall, 41.4\% of participants described their chatbot relationship in ways that aligned with multiple usage categories, underscoring the multifaceted nature of chatbot interactions.

Evidence from donated chat transcripts shows a similar pattern. Among participants who shared chat histories, 92.9\% included at least one conversation classified as companionship-oriented. Even among those who did not identify companionship as their primary usage in the survey, 47.8\% still had at least one conversation exhibiting companion-like engagement. 
Relational elements also appeared frequently among users who reported entertainment as their primary chatbot usage. In this group, 52.69\% described their chatbot relationship using relational language such as ``friend'' or ``emotional support''. Consistent patterns appear in their donated conversations: 80.13\% included emotional support interactions and 78.85\% included romantic exploration.
Taken together, these findings indicate that chatbot engagement often combines relational and non-relational motivations. Rather than fitting neatly into a single functional category, many interactions incorporate overlapping purposes. In some cases, companion-like engagement can arise in role-play contexts, including romantic or sexual scenarios, where companionship-like interaction functions as part of narrative engagement rather than ongoing relational maintenance. At the same time, sustained interaction with conversational agents may encourage patterns of communication resembling human companionship. These dynamics raise questions about how repeated engagement with AI partners could shape expectations of intimacy, responsiveness, and emotional reciprocity in human relationships.

\paragraph{Users engage with chatbots on a range of intimate and sensitive topics.} 
Looking further into the chat content, we find that users converse with chatbots on a range of intimate and potentially risky topics. As shown in Figure~\ref{fig: usage_relationship}(b), the most prevalent conversation topic was seeking \emph{emotional and social support}, such as discussing personal challenges and health-related concerns, which appeared in $80.3\%$ of donated chat sessions. We observed \emph{romantic and intimacy roleplay} in $68.0\%$ of sessions, suggesting that affective bonds and imaginative forms of intimacy arise in chatbot interactions. In addition, $30.7\%$ of sessions included  \emph{risky and dark roleplay} involving illicit or taboo scenarios.

In summary, companionship-oriented engagement appears more prevalent in practice than explicit self-reports suggest. These interactions often involved emotionally sensitive domains, including support-seeking, romantic and sexual roleplay, and in some cases even risky or toxic scenarios. These patterns indicate that companionship-like dynamics are both more prevalent and more complex than users' stated primary motives suggest, as relational engagement frequently overlaps with other forms of interaction.

\subsection{Users with less human social support are more likely to seek chatbot companionship} \label{section: human_social_support_companionship}

First, we examine whether engagement with chatbots differs across users' broader social environments. The \emph{Social Compensation hypothesis} posits that individuals with limited human social networks may turn to alternative sources of connection, such as chatbots, to fulfill unmet emotional needs~\cite{kraut2002internet, teppers2014loneliness, lee2013lonely, tian2013social, weidman2012compensatory, hood2018loneliness}.
From this view, AI companionship serves a beneficial role, providing a low-cost outlet for emotional expression and social fulfillment, particularly among socially isolated users. Drawing on this social theory, we test whether users with smaller social networks are more inclined toward companionship-type chatbot engagement.

Table~\ref{table: predict_interaction_aligned} presents regression results examining the influence of social network scale on various measures of chatbot interaction. Results show no evidence that social network scale was associated with overall chatbot intensity.
However, social network scale was negatively associated with companionship as a self-reported motivation ($\beta = -0.03$, $p < .001$),
indicating that users with smaller social networks are more likely to turn to chatbots for companionship. 
Furthermore, social network scale was negatively associated with self-disclosure ($\beta = -0.10$, $p < .001$). This suggests that individuals with fewer social connections are more inclined to share personal information with chatbots.

\begin{table}[htbp]
\centering
\large
\resizebox{\textwidth}{!}{%
\begin{tabular}{l c c c c c}
\toprule
 &
\textbf{(1) Intensity} &
\textbf{(2) Companionship\textsubscript{Prim.}} &
\textbf{(3) Companionship\textsubscript{Desc.}} &
\textbf{(4) Companionship\textsubscript{Chat}} &
\textbf{(5) Self-Disclosure} \\
\midrule

Intercept &
\makecell{%
$-0.49$\\
$[-0.68,\ -0.30]$\\
$p < 0.001$
} &
\makecell{%
$0.14$\\
$[0.07,\ 0.20]$\\
$p < 0.001$
} &
\makecell{%
$0.52$\\
$[0.42,\ 0.63]$\\
$p < 0.001$
} &
\makecell{%
$0.68$\\
$[-1.86	,\ 3.22]$\\
$p = 0.600$
} &
\makecell{%
$-0.17$\\
$[-0.37,\ 0.03]$\\
$p = 0.094$
} \\
\addlinespace[8pt]

Social Network &
\makecell{%
$0.01$\\
$[-0.04,\ 0.07]$\\
$p = 0.672$
} &
\makecell{%
$-0.03$\\
$[-0.05,\ -0.01]$\\
$p < 0.001$
} &
\makecell{%
$-0.03$\\
$[-0.06,\ 0.00]$\\
$p = 0.056$
} &
\makecell{%
$-0.00$\\
$[-0.12,\ 0.12]$\\
$p = 0.970$
} &
\makecell{%
$-0.10$\\
$[-0.16,\ -0.04]$\\
$p < 0.001$
} \\
\addlinespace[8pt]

Tenure &
\makecell{%
$0.25$\\
$[0.20,\ 0.31]$\\
$p < 0.001$
} &
\makecell{%
$0.02$\\
$[-0.00,\ 0.03]$\\
$p = 0.091$
} &
\makecell{%
$0.02$\\
$[-0.01,\ 0.04]$\\
$p = 0.310$
} &
\makecell{%
$-0.02$\\
$[-0.20,\ 0.16]$\\
$p = 0.809$
} &
\makecell{%
$0.13$\\
$[0.07,\ 0.19]$\\
$p < 0.001$
} \\
\addlinespace[8pt]

Male &
\makecell{%
$0.09$\\
$[-0.02,\ 0.21]$\\
$p = 0.099$
} &
\makecell{%
$0.01$\\
$[-0.03,\ 0.05]$\\
$p = 0.680$
} &
\makecell{%
$0.02$\\
$[-0.04,\ 0.08]$\\
$p = 0.615$
} &
\makecell{%
$-0.25$\\
$[-0.57,\ 0.07]$\\
$p = 0.121$
} &
\makecell{%
$0.07$\\
$[-0.05,\ 0.19]$\\
$p = 0.266$
} \\
\addlinespace[8pt]

Non-binary &
\makecell{%
$-0.38$\\
$[-0.64,\ -0.13]$\\
$p = 0.003$
} &
\makecell{%
$0.18$\\
$[0.10,\ 0.27]$\\
$p < 0.001$
} &
\makecell{%
$0.03$\\
$[-0.11,\ 0.17]$\\
$p = 0.677$
} &
\makecell{%
$-0.01$\\
$[-0.85,\ 0.83]$\\
$p = 0.979$
} &
\makecell{%
$-0.06$\\
$[-0.33,\ 0.21]$\\
$p = 0.679$
} \\
\addlinespace[8pt]

Age &
\makecell{%
$0.02$\\
$[0.01,\ 0.02]$\\
$p < 0.001$
} &
\makecell{%
$-0.00$\\
$[-0.00,\ 0.00]$\\
$p = 0.604$
} &
\makecell{%
$-0.00$\\
$[-0.00,\ 0.00]$\\
$p = 0.748$
} &
\makecell{%
$-0.02$\\
$[-0.05,\ 0.01]$\\
$p = 0.128$
} &
\makecell{%
$0.01$\\
$[0.00,\ 0.01]$\\
$p = 0.031$
} \\
\addlinespace[8pt]

Single &
\makecell{%
$-0.20$\\
$[-0.31,\ -0.08]$\\
$p < 0.001$
} &
\makecell{%
$-0.05$\\
$[-0.08,\ -0.01]$\\
$p = 0.022$
} &
\makecell{%
$-0.02$\\
$[-0.08,\ 0.04]$\\
$p = 0.537$
} &
\makecell{%
$0.04$\\
$[-0.55,\ 0.63]$\\
$p = 0.892$
} &
\makecell{%
$-0.09$\\
$[-0.21,\ 0.03]$\\
$p = 0.135$
} \\

\midrule
$R^2$ &
$0.135$ & $0.032$ & $0.005$ & --- & $0.033$ \\\addlinespace[8pt]
Adjusted $R^2$ &
$0.130$ & $0.027$ & $-0.001$ & --- & $0.028$ \\\addlinespace[8pt]
$N$ &
$1131$ & $1131$ & $1131$ & $1131$ & $1131$ \\\addlinespace[8pt]
df &
$1124$ & $1124$ & $1124$ & $1116$ & $1124$ \\

\bottomrule
\end{tabular}
}
\\[1em]
\caption{
\textbf{Regression models estimating associations between demographic characteristics, social network scale, and chatbot interaction measures.}
Values are standardized coefficients ($\beta$) with 95\% confidence intervals and two-sided \textit{p}-values from \textit{t}-tests.
Model (1) estimates interaction intensity. Model (2) estimates self-reported primary companionship usage. Model (3) estimates companionship based on open-ended relationship descriptions. Model (4) estimates the proportion of companionship-related sessions in chat history. Model (5) estimates self-reported self-disclosure in chatbot interactions. 
}
\label{table: predict_interaction_aligned}
\end{table}

Beyond social network size, other individual differences were associated with chatbot engagement.
Users who identify as non-binary had lower interaction intensity overall ($\beta = -0.38$, $p = .003$) but higher companionship usage ($\beta = 0.18$, $p < .001$). 
These findings are also in line with the Social Compensation hypothesis, which suggests that individuals in marginalized gender groups may selectively seek emotional connection through chatbots when offline support is limited.
Surprisingly, users who identified themselves as single reported both lower overall interaction intensity ($\beta = -0.20$, $p < .001$) and companionship motivation ($\beta = -0.05$, $p = .022$). This pattern contrasts with the Social Compensation hypothesis. However, the modest correlation between single status and social network size ($r = -0.11$) suggests that being single does not necessarily indicate limited social support, as many single individuals may still have strong offline networks, reducing the need for chatbot companionship.

Finally, tenure of chatbot use was positively associated with interaction intensity ($\beta = 0.25$, $p < .001$), indicating that users with longer usage histories tend to engage more intensively. 
This pattern is consistent with a potential reinforcement cycle, where continued use deepens engagement, which in turn may encourage further use, suggesting how chatbot relationships can become more embedded over time.

\subsection{While more overall chatbot use is associated with higher well-being, more companionship use is associated with lower well-being} \label{section: intensity_companionship}
We next examine how different forms of chatbot interaction relate to psychological well-being, focusing on two dimensions: overall interaction intensity and usage type (companionship vs. non-companionship). We draw on both survey and behavioral data. 

More intense chatbot use was significantly associated with higher psychological well-being. As shown in Table~\ref{table: full_combined_model_results}, in both Model~(1), based on participants' self-reported primary purpose, and Model~(2), using relationship descriptions, higher intensity was significantly associated with greater well-being (Model 1: $\beta = 0.27$, $p < .001$; Model 2: $\beta = 0.29$, $p < .001$) . These findings suggest that individuals who engage with chatbots more frequently, share personal matters with a larger number of chatbots, and incorporate these interactions into their daily routines tend to report higher well-being. This positive association was not observed in Model~(3) using chat history data  ($\beta = -0.08$, $p = .431$).

\sisetup{
  table-format=3.2,
  detect-weight=true,
  detect-family=true,
  input-symbols = {***, **, *},
  table-space-text-post = ***
}

\begin{table}[htbp]
\centering
\large
\resizebox{\textwidth}{!}{%
\begin{tabular}{l c c c | c c c | c c}
\toprule
\multicolumn{9}{c}{\textbf{Dependent Variable: Well-being}} \\
\midrule
&
\multicolumn{3}{c|}{\textbf{(1) Companionship\textsubscript{Prim.}}\textsuperscript{\S}} &
\multicolumn{3}{c|}{\textbf{(2) Companionship\textsubscript{Desc.}}\textsuperscript{\S}} &
\multicolumn{2}{c}{\textbf{(3) Companionship\textsubscript{Chat}}\textsuperscript{\S}} \\
\midrule

Intercept
& \makecell{$4.57$\\$[4.31,\ 4.83]$\\$p < 0.001$}
& \makecell{$4.57$\\$[4.31,\ 4.83]$\\$p < 0.001$}
& \makecell{$4.43$\\$[4.17,\ 4.69]$\\$p < 0.001$}
& \makecell{$4.68$\\$[4.41,\ 4.95]$\\$p < 0.001$}
& \makecell{$4.68$\\$[4.41,\ 4.95]$\\$p < 0.001$}
& \makecell{$4.52$\\$[4.24,\ 4.79]$\\$p < 0.001$}
& \makecell{$3.66$\\$[0.61,\ 6.71]$\\$p = 0.019$}
& \makecell{$3.63$\\$[0.73,\ 6.52]$\\$p = 0.014$}
\\ \addlinespace[8pt]

Intensity
& \makecell{$0.27$\\$[0.19,\ 0.35]$\\$p < 0.001$}
& \makecell{$0.30$\\$[0.22,\ 0.38]$\\$p < 0.001$}
& ---
& \makecell{$0.29$\\$[0.21,\ 0.37]$\\$p < 0.001$}
& \makecell{$0.29$\\$[0.18,\ 0.39]$\\$p < 0.001$}
& ---
& \makecell{$-0.08$\\$[-0.27,\ 0.12]$\\$p = 0.431$}
& \makecell{$-0.08$\\$[-0.27,\ 0.11]$\\$p = 0.415$}
\\ \addlinespace[8pt]

\textit{Companionship}\textsuperscript{\S}
& \makecell{$-0.48$\\$[-0.70,\ -0.25]$\\$p < 0.001$}
& \makecell{$-0.40$\\$[-0.64,\ -0.17]$\\$p < 0.001$}
& \makecell{$-0.28$\\$[-0.52,\ -0.03]$\\$p = 0.030$}
& \makecell{$-0.32$\\$[-0.47,\ -0.17]$\\$p < 0.001$}
& \makecell{$-0.32$\\$[-0.47,\ -0.17]$\\$p < 0.001$}
& \makecell{$-0.22$\\$[-0.38,\ -0.07]$\\$p = 0.005$}
& \makecell{$-0.27$\\$[-0.46,\ -0.09]$\\$p = 0.004$}
& \makecell{$-0.32$\\$[-0.52,\ -0.13]$\\$p = 0.001$}
\\ \addlinespace[8pt]

Intensity $\times$ \textit{Companionship}\textsuperscript{\S}
& ---
& \makecell{$-0.31$\\$[-0.56,\ -0.06]$\\$p = 0.014$}
& ---
& ---
& \makecell{$0.01$\\$[-0.14,\ 0.16]$\\$p = 0.934$}
& ---
& ---
& \makecell{$-0.13$\\$[-0.32,\ 0.05]$\\$p = 0.144$}
\\ \addlinespace[8pt]

Self-disclosure
& ---
& ---
& \makecell{$0.08$\\$[0.00,\ 0.16]$\\$p = 0.049$}
& ---
& ---
& \makecell{$0.10$\\$[-0.01,\ 0.20]$\\$p = 0.063$}
& ---
& ---
\\ \addlinespace[8pt]

\textit{Companionship}\textsuperscript{\S} $\times$ Self-disclosure
& ---
& ---
& \makecell{$-0.38$\\$[-0.63,\ -0.14]$\\$p = 0.002$}
& ---
& ---
& \makecell{$-0.11$\\$[-0.26,\ 0.05]$\\$p = 0.185$}
& ---
& ---
\\ \addlinespace[8pt]

Tenure
& \makecell{$0.01$\\$[-0.07,\ 0.08]$\\$p = 0.881$}
& \makecell{$0.00$\\$[-0.07,\ 0.08]$\\$p = 0.959$}
& \makecell{$0.06$\\$[-0.01,\ 0.14]$\\$p = 0.106$}
& \makecell{$-0.00$\\$[-0.08,\ 0.07]$\\$p = 0.951$}
& \makecell{$-0.00$\\$[-0.08,\ 0.07]$\\$p = 0.952$}
& \makecell{$0.06$\\$[-0.01,\ 0.14]$\\$p = 0.096$}
& \makecell{$-0.05$\\$[-0.30,\ 0.19]$\\$p = 0.668$}
& \makecell{$-0.05$\\$[-0.29,\ 0.19]$\\$p = 0.671$}
\\ \addlinespace[8pt]

Male
& \makecell{$0.17$\\$[0.02,\ 0.32]$\\$p = 0.023$}
& \makecell{$0.16$\\$[0.01,\ 0.31]$\\$p = 0.036$}
& \makecell{$0.19$\\$[0.04,\ 0.34]$\\$p = 0.013$}
& \makecell{$0.17$\\$[0.02,\ 0.32]$\\$p = 0.023$}
& \makecell{$0.17$\\$[0.02,\ 0.32]$\\$p = 0.023$}
& \makecell{$0.19$\\$[0.04, 0.35]$\\$p = 0.013$}
& \makecell{$0.16$\\$[-0.28,\ 0.60]$\\$p = 0.476$}
& \makecell{$0.17$\\$[-0.27,\ 0.61]$\\$p = 0.444$}
\\ \addlinespace[8pt]

Non-binary
& \makecell{$-0.33$\\$[-0.67,\ 0.02]$\\$p = 0.062$}
& \makecell{$-0.31$\\$[-0.65,\ 0.03]$\\$p = 0.078$}
& \makecell{$-0.39$\\$[-0.74,\ -0.05]$\\$p = 0.026$}
& \makecell{$-0.40$\\$[-0.74,\ -0.06]$\\$p = 0.023$}
& \makecell{$-0.40$\\$[-0.74,\ -0.06]$\\$p = 0.023$}
& \makecell{$-0.49$\\$[-0.84,\ -0.15]$\\$p = 0.005$}
& \makecell{$0.12$\\$[-0.94,\ 1.18]$\\$p = 0.821$}
& \makecell{$0.16$\\$[-0.86,\ 1.19]$\\$p = 0.755$}
\\ \addlinespace[8pt]

Age
& \makecell{$0.01$\\$[0.01,\ 0.02]$\\$p < 0.001$}
& \makecell{$0.01$\\$[0.01,\ 0.02]$\\$p < 0.001$}
& \makecell{$0.02$\\$[0.01,\ 0.02]$\\$p < 0.001$}
& \makecell{$0.01$\\$[0.01,\ 0.02]$\\$p < 0.001$}
& \makecell{$0.01$\\$[0.01,\ 0.02]$\\$p < 0.001$}
& \makecell{$0.02$\\$[0.01,\ 0.02]$\\$p < 0.001$}
& \makecell{$0.01$\\$[-0.02,\ 0.05]$\\$p = 0.396$}
& \makecell{$0.01$\\$[-0.02,\ 0.05]$\\$p = 0.426$}
\\ \addlinespace[8pt]

Single
& \makecell{$-0.47$\\$[-0.62,\ -0.31]$\\$p < 0.001$}
& \makecell{$-0.46$\\$[-0.61,\ -0.31]$\\$p < 0.001$}
& \makecell{$-0.51$\\$[-0.66, -0.35]$\\$p < 0.001$}
& \makecell{$-0.45$\\$[-0.60,\ -0.29]$\\$p < 0.001$}
& \makecell{$-0.45$\\$[-0.60,\ -0.29]$\\$p < 0.001$}
& \makecell{$-0.50$\\$[-0.66,\ -0.35]$\\$p < 0.001$}
& \makecell{$-0.24$\\$[-0.97,\ 0.48]$\\$p = 0.510$}
& \makecell{$-0.22$\\$[-0.92,\ 0.48]$\\$p = 0.533$}
\\

\midrule
$R^2$
& $0.136$ & $0.140$ & $0.109$
& $0.136$ & $0.136$ & $0.100$
& --- & --- \\
\addlinespace[8pt]

Adjusted $R^2$
& $0.130$ & $0.134$ & $0.103$
& $0.131$ & $0.130$ & $0.094$
& --- & --- \\
\addlinespace[8pt]

$N$
& $1131$ & $1131$ & $1131$
& $1131$ & $1131$ & $1131$
& $1131$ & $1131$ \\
\addlinespace[8pt]

df
& $1123$ & $1122$ & $1122$
& $1123$ & $1122$ & $1122$
& $1115$ & $1114$ \\
\bottomrule
\end{tabular}
}
\caption{
\textbf{Regression models examining associations of chatbot companionship, interaction intensity, and self-disclosure with well-being.} Values are standardized coefficients ($\beta$) with 95\% confidence intervals and two-sided \textit{p}-values from \textit{t}-tests. Companionship is operationalized through (1) single-choice main purpose ($\textit{Companionship}_{\textit{Prim.}}$), (2) classified relationship descriptions ($\textit{Companionship}_{\textit{Desc.}}$), and (3) proportion of companionship-related messages in chat history ($\textit{Companionship}_{\textit{Chat}}$). For the first two types ($\textit{Companionship}_{\textit{Prim.}}$ and $\textit{Companionship}_{\textit{Desc.}}$), both intensity and self-disclosure interactions are included. For the third ($\textit{Companionship}_{\textit{Chat}}$), only intensity interaction was estimated due to non-convergence of the self-disclosure model.}
\label{table: full_combined_model_results}
\end{table}

However, holding constant overall chatbot intensity, companionship-oriented chatbot use was consistently associated with lower psychological well-being across all three models. 
In Model~(1), participants who self-reported using chatbots primarily for companionship reported significantly lower well-being compared to those with other motivations ($\beta = -0.48$, $p < .001$). 
Model~(2), which classified companionship based on relationship descriptions, showed a similarly negative association ($\beta = -0.32$, $p < .001$). Model~(3), which relied on the proportion of companionship-oriented sessions in chat histories, also revealed significant associations ($\beta = -0.27$, $p = .004$).
Overall, these consistent patterns indicate that companionship use, whether self-reported or inferred from behavioral data, is associated with lower well-being. %
These results also parallel findings in adjacent domains such as social media, television, and gaming, where well-being outcomes are often shaped less by overall use than by patterns of use and interaction~\cite{verduyn2017social, ellison2007benefits, roberts2023instagram}. %

\subsection{The negative relationship between companionship use and well-being is stronger when chatbot use is more intense}  \label{section: intensity_moderation}
\begin{figure}[tb]
\centering
\includegraphics[width=\textwidth]{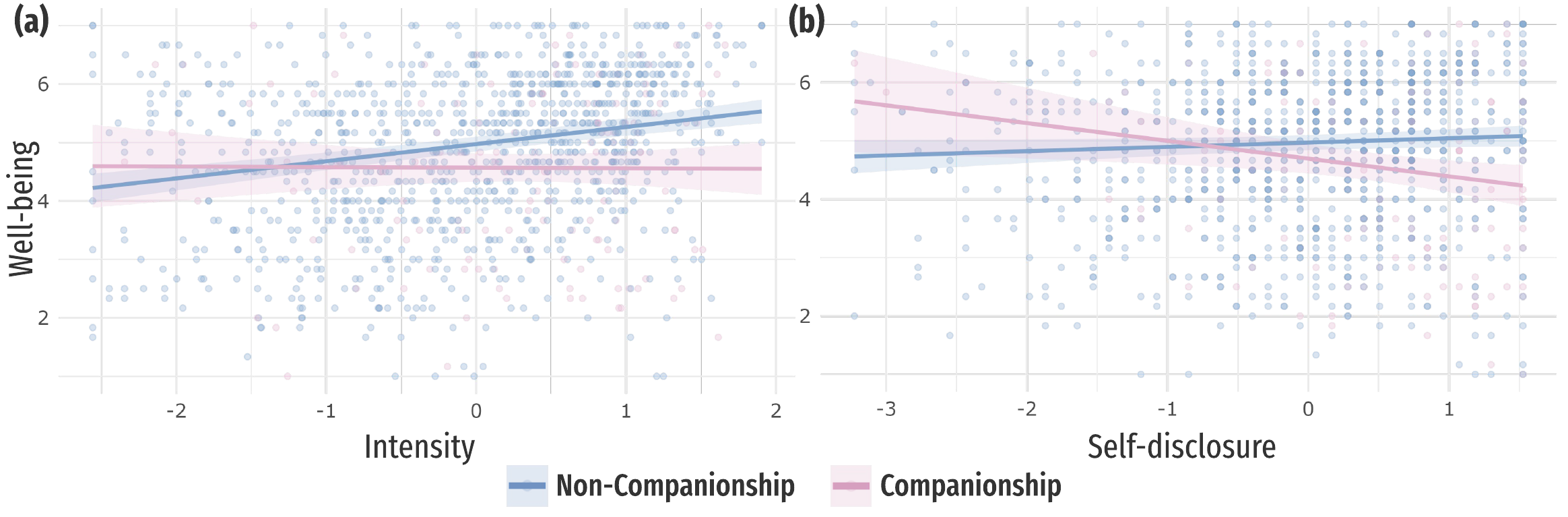}
\caption{\textbf{Interaction between companionship use and other interaction measures in predicting well-being.}
Figures present results from two-sided regression models examining associations between companionship chatbot use and well-being, including interaction terms with (a) interaction intensity and (b) self-disclosure.
In Figure (a), the interaction between companionship use and intensity was statistically significant 
($\beta = -0.31$, 95\% CI [$-0.56$, $-0.06$]; $t = -2.47$, df = $1122$, $p = 0.014$). 
In Figure (b), the interaction between companionship use and self-disclosure was also statistically significant 
($\beta = -0.38$, 95\% CI [$-0.63$, $-0.14$]; $t = -3.08$, df = $1122$, $p = 0.002$). 
Solid lines indicate model-predicted mean well-being, and shaded bands represent 95\% confidence intervals for the mean prediction. Points represent individual observed data.}
\label{fig: intensity_companion_result}
\end{figure}

Given that companionship use was associated with lower well-being, we next examined whether the intensity of use moderated the relationship between companionship use and well-being. We include an interaction term between companionship chatbot use and interaction intensity in regression models examining psychological well-being.
The results reported in Table~\ref{table: full_combined_model_results} revealed a negative interaction between companionship and chatbot intensity in the self-reported primary usage model ($\beta = -0.31$, $p = .014$; see Table~\ref{table: full_combined_model_results} and Figure~\ref{fig: intensity_companion_result}(a)). This interaction indicates that the negative association between companionship use of chatbots and well-being was stronger among users who used the chatbots more intensively.
One explanation is that when people use chatbots for companionship, relying on them more intensely may build habits of turning to chatbots instead of people, which contributes to lower well-being. Another explanation is that people with lower well-being may be more likely both to seek companionship from chatbots and to use them more intensely. In either case, it points to a vulnerable subgroup for whom chatbot reliance may serve as a coping strategy for relational needs.

\subsection{Companionship use is associated with lower well-being at higher levels of self-disclosure} \label{section: self-disclosure}

Self-disclosure is a well-established predictor of intimacy and psychological well-being in human-human relationships~\cite{bazarova2012public, utz2015function, deters2013does, luo2020selfdisclosure}. 
For AI companionship, users may feel especially comfortable disclosing sensitive information to chatbots due to perceptions of anonymity and free from social judgment~\cite{ta2020user, skjuve2021my, bickmore2001relational}. While such disclosure can offer emotional relief and simulate intimacy~\cite{jiang2011disclosure, bazarova2012public}, it also carries risks, especially in digital settings that lack clear boundaries or protective social norms. Concerns include emotional overexposure, privacy vulnerabilities, and attachment to chatbots in ways that may not reflect socially regulated or psychologically safe interactions~\cite{vogel2003seek, chen2017antecedents, lin2012sharing, gil2015facebook, reece2017instagram}.
To assess these relationships in human-chatbot interactions, we investigate how users' willingness to disclose personal information to chatbots relates to their well-being.

When controlling for companionship use, self-disclosure was positively associated with well-being ($\beta= 0.08$, $p =.049$) in the self-reported primary usage model and there is no significant association was observed in the other two models.
However, as shown in Table~\ref{table: full_combined_model_results}  and Figure~\ref{fig: intensity_companion_result}(b), there was a significant negative interaction between companionship usage and self-disclosure in the self-reported usage model ($\beta = -0.38$, $p = .002$). This result indicates that the negative association between companionship use of chatbots and well-being is stronger when people self-disclose more to them. 
Interaction effects were not significant in the other model.
Alternatively, another possibility is that individuals with lower well-being may be more likely to seek companionship through chatbots, and their higher level of self-disclosure may intensify this tendency. In this way, the disclosure in companionship contexts may indicate a heightened vulnerability, where people reveal more personal and sensitive information without the reciprocity exchange typical of human relationships and with greater exposure to privacy risks.

\begin{figure}[!htbp]
  \centering
  \includegraphics[width=\textwidth]{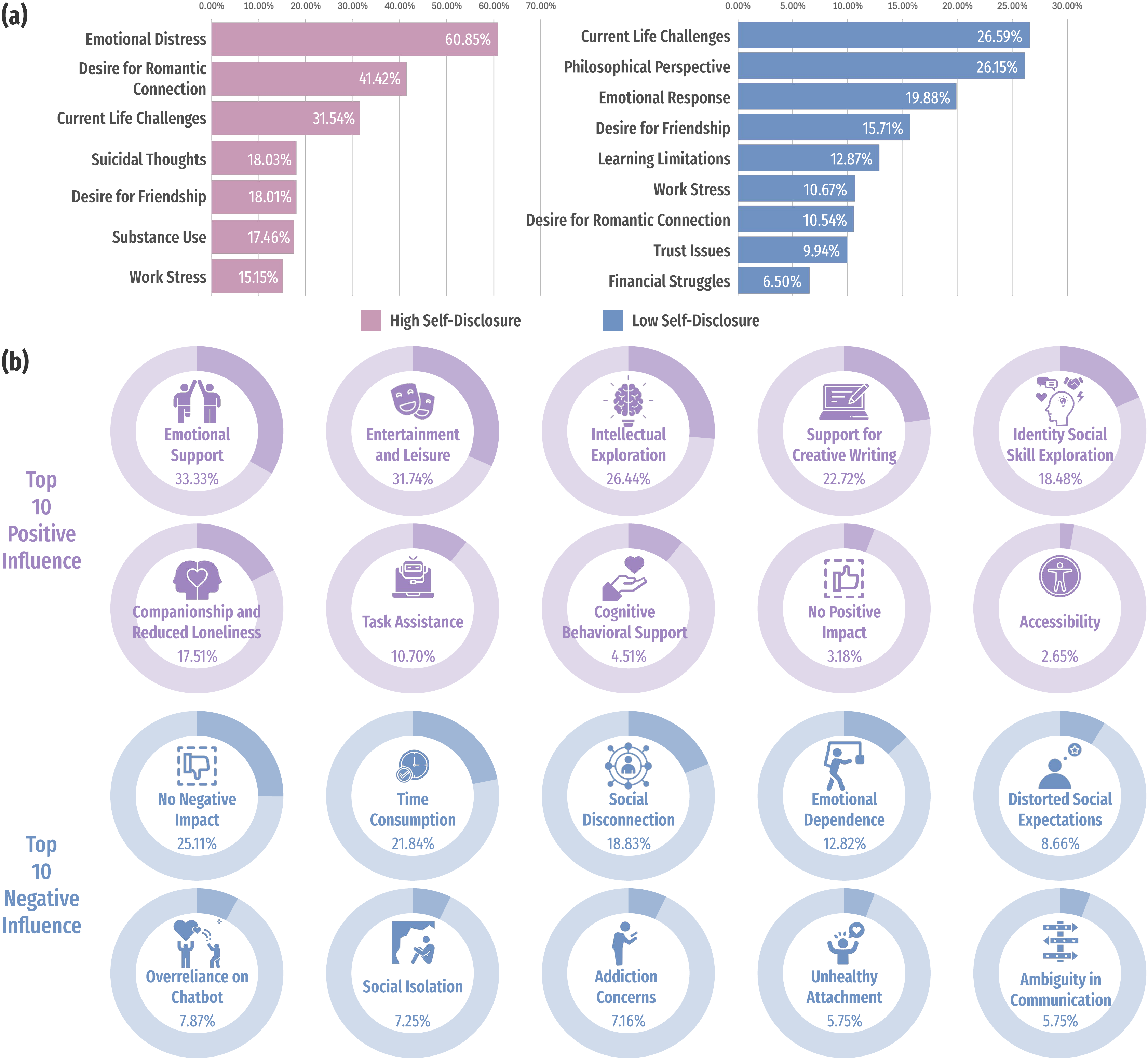}
  \caption{\textbf{Topic modeling analysis of user-donated chat histories and self-reported reflections on chatbot interactions.} All themes were derived using TopicGPT (see the Topic modeling section in Supplementary Information). %
  (a) Self-disclosure topics classified as high (pink) or low (blue) based on the criteria from \citet{balani2015detecting}.
  (b) User-perceived positive and negative influences of chatbot interaction.}
  \label{fig: content_analysis_result}
\end{figure}

To better understand what users disclosed to chatbots, we applied LLM-based topic modeling to chat histories, using disclosure-level definitions developed by \citet{balani2015detecting}. Both high- and low-disclosure conversations covered some similar broad themes, such as life challenges, desire for friendship, and desire for romantic relationships, but they reflected different levels of emotional vulnerability and depth. Low self-disclosure conversations tended to stay at a surface level, including brief mentions of daily routines, general dating intentions, or abstract reflections on personal growth. In contrast, high self-disclosure conversations often contained personal and emotionally intense stories. For example, some users described persistent anxiety about death without knowing where these feelings came from, or shared long-term experiences of mental health issues linked to childhood trauma. Others expressed feelings of rejection and a longing for care, such as recalling being called ``bad'' by foster parents and anxiously asking if the chatbot could become a ``pretend parent''. 
Conversations characterized by high self-disclosure predominantly involved emotionally charged and sensitive content, including emotional distress (60.8\%), suicidal thoughts (18.0\%), and substance use (17.5\%). 
In contrast, low self-disclosure conversations addressed less personal and more abstract topics, such as philosophical perspectives (26.2\%), learning limitations (12.9\%), and general issues like work stress or trust (see Figure~\ref{fig: content_analysis_result}(a)). 
As observed, high self-disclosure involves not only emotionally intense content, such as personal struggles and distress, but also highly sensitive and risky topics, including suicidal thoughts. This pattern suggests that disclosure in chatbot interactions may not confer the same psychological benefits observed in human relationships, and may instead coincide with greater emotional vulnerability and potential risks.

\subsection{The relationship between AI chatbot usage and well-being is correlated with users' offline social support} \label{section: human_social_support_moderation}
Finally, we examine how the association between chatbot use and well-being differs based on users' level of offline social support. As shown in Table~\ref {table: combined_wellbeing_models}, we find that having a strong offline social network was positively associated with well-being, which mirrors the established research on the psychological benefits of human connection~\cite{morina2021potential, holt2024social}. Since chatbots provide a new type of social connection, we now turn to how different patterns of chatbot use interact with the benefits of offline social support.

\begin{table}[htbp]
\centering
\normalsize
\resizebox{\textwidth}{!}{%
\begin{tabular}{l c c c c}
\toprule
\multicolumn{5}{c}{\textbf{Dependent Variable: Well-being}} \\
\midrule
&
\textbf{(1) Intensity} &
\textbf{(2) Companionship\textsubscript{Prim.}}\textsuperscript{\S} &
\textbf{(3) Companionship\textsubscript{Desc.}}\textsuperscript{\S} &
\textbf{(4) Companionship\textsubscript{Chat}}\textsuperscript{\S} \\
\midrule

Intercept
& \makecell{$4.53$\\$[4.29,\ 4.76]$\\$p < 0.001$}
& \makecell{$4.46$\\$[4.22,\ 4.70]$\\$p < 0.001$}
& \makecell{$4.50$\\$[4.25,\ 4.75]$\\$p < 0.001$}
& \makecell{$3.94$\\$[0.53,\ 7.36]$\\$p = 0.024$} \\
\addlinespace[8pt]

\textit{Companionship}\textsuperscript{\S}
& ---
& \makecell{$-0.22$\\$[-0.44,\ -0.00]$\\$p = 0.046$}
& \makecell{$-0.14$\\$[-0.28,\ -0.00]$\\$p = 0.043$}
& \makecell{$-0.27$\\$[-0.43,\ -0.11]$\\$p < 0.001$} \\
\addlinespace[8pt]

\textit{Intensity}
& \makecell{$0.25$\\$[0.18,\ 0.32]$\\$p < 0.001$}
& --- & --- & --- \\
\addlinespace[8pt]

Social network
& \makecell{$0.51$\\$[0.44,\ 0.57]$\\$p < 0.001$}
& \makecell{$0.49$\\$[0.41,\ 0.56]$\\$p = < 0.001$}
& \makecell{$0.49$\\$[0.39,\ 0.59]$\\$p < 0.001$}
& \makecell{$0.76$\\$[0.62,\ 0.91]$\\$p < 0.001$} \\
\addlinespace[8pt]

\textit{Companionship}\textsuperscript{\S} $\times$ Social Network
& ---
& \makecell{$0.07$\\$[-0.14,\ 0.28]$\\$p = 0.514$}
& \makecell{$0.01$\\$[-0.12,\ 0.15]$\\$p = 0.838$}
& \makecell{$-0.02$\\$[-0.16,\ 0.13]$\\$p = 0.836$} \\
\addlinespace[8pt]

\textit{Intensity} $\times$ Social Network
& \makecell{$-0.11$\\$[-0.18,\ -0.05]$\\$p < 0.001$}
& --- & --- & --- \\
\addlinespace[8pt]

Tenure
& \makecell{$-0.02$\\$[-0.08,\ 0.05]$\\$p = 0.668$}
& \makecell{$0.05$\\$[-0.02,\ 0.12]$\\$p = 0.147$}
& \makecell{$0.05$\\$[-0.02,\ 0.12]$\\$p = 0.162$}
& \makecell{$-0.08$\\$[-0.31,\ 0.15]$\\$p = 0.492$} \\
\addlinespace[8pt]

Male
& \makecell{$0.09$\\$[-0.05,\ 0.23]$\\$p = 0.193$}
& \makecell{$0.11$\\$[-0.03,\ 0.25]$\\$p = 0.128$}
& \makecell{$0.11$\\$[-0.03,\ 0.25]$\\$p = 0.132$}
& \makecell{$0.01$\\$[-0.40,\ 0.41]$\\$p = 0.973$} \\
\addlinespace[8pt]

Non-binary
& \makecell{$-0.49$\\$[-0.80,\ -0.17]$\\$p = 0.002$}
& \makecell{$-0.53$\\$[-0.85,\ -0.21]$\\$p = 0.001$}
& \makecell{$-0.57$\\$[-0.89,\ -0.25]$\\$p < 0.001$}
& \makecell{$-0.27$\\$[-1.36,\ 0.83]$\\$p = 0.633$} \\
\addlinespace[8pt]

Age
& \makecell{$0.01$\\$[0.01,\ 0.02]$\\$p < 0.001$}
& \makecell{$0.02$\\$[0.01,\ 0.02]$\\$p < 0.001$}
& \makecell{$0.02$\\$[0.01,\ 0.02]$\\$p < 0.001$}
& \makecell{$0.02$\\$[-0.02,\ 0.05]$\\$p = 0.387$} \\
\addlinespace[8pt]

Single
& \makecell{$-0.34$\\$[-0.48,\ -0.20]$\\$p < 0.001$}
& \makecell{$-0.41$\\$[-0.55,\ -0.27]$\\$p < 0.001$}
& \makecell{$-0.40$\\$[-0.55,\ -0.26]$\\$p < 0.001$}
& \makecell{$-0.17$\\$[-0.94,\ 0.61]$\\$p = 0.676$} \\

\midrule
$R^2$
& $0.270$ & $0.236$ & $0.235$ & --- \\
\addlinespace[6pt]
Adjusted $R^2$
& $0.265$ & $0.230$ & $0.230$ & --- \\
\addlinespace[6pt]
$N$
& $1131$ & $1131$ & $1131$ & $1131$ \\
\addlinespace[6pt]
df
& $1122$ & $1122$ & $1122$ & $1114$ \\
\bottomrule
\end{tabular}
}
\caption{\textbf{Regression models examining associations of chatbot companionship and interaction intensity with well-being.} Values are standardized coefficients ($\beta$) with 95\% confidence intervals and two-sided \textit{p}-values from \textit{t}-tests. Model 1 tests the interaction between social network scale and chatbot interaction intensity. Models 2-4 test the interaction between social network scale and three measures of companionship: self-reported primary usage, relationship descriptions, and the proportion of companionship-related messages.}
\label{table: combined_wellbeing_models}
\end{table}

From Sec.~\ref{section: human_social_support_companionship}, we already observed that users with less offline support were more likely to engage with chatbots as companions, but is this interaction beneficial for their well-being? Following the Social Compensation hypothesis, we might expect that chatbot companions provide a positive alternative for individuals with smaller close social networks that improves psychological well-being. 
However, as shown in Table~\ref{table: combined_wellbeing_models}, we did not observe a statistically significant interaction between social network scale and chatbot companionship ($\beta = 0.07$, $p = .51$). Bayesian model comparison further indicated that the data were more consistent with the model without the interaction than with the interaction model ($BF_{01} = 6.55$).
An equivalence test using $|\beta| = .10$ as the smallest effect size of interest \cite{godard2024active} further indicated that the interaction fell within the predefined equivalence bounds, $t(1122) = -3.15$, $p < .001$.
These results indicate no evidence of a meaningful moderation effect of companionship-oriented chatbot use on the association between offline social network size and well-being.
As discussed in Sec.~\ref{section: intensity_companionship}, we see that overall, companionship usage is negatively correlated with well-being. Thus, while users with less social support are more likely to seek companionship through chatbots, these interactions do not appear to offset the association between low social support and reduced well-being. 

In addition, we observed a significant negative interaction between chatbot use intensity and offline social network scale ($\beta = -0.11$, $p < .001$), indicating that the positive association between offline social networks and well-being was weaker at higher levels of chatbot-use intensity. This pattern is aligned with \emph{Social Substitution hypothesis}, which posits that heavy engagement with chatbot may weaken the psychological benefits typically derived from offline social relationships, without necessarily improving overall well-being.

Overall, our findings point to the relational limits of chatbot use: \textbf{while chatbots may augment existing social networks, they do not effectively substitute them. The influence of AI companionship depends not only on how it is used but also on the social environment in which it is embedded.}

\section{Discussion}\label{discussion}
This study offers a broad investigation into the psychological implications of AI companionship, drawing on self-reported and behavioral data from a sample of active Character.AI users. Our analysis shows that using AI chatbots for companionship is prevalent among our surveyed users. These chatbots are becoming increasingly integrated into users' emotional and social lives, functioning not only as tools but also as relational partners that assume the role of companions, friends, and even romantic partners. Nonetheless, the findings reveal that the relationship between AI companionship and well-being is complex. Rather than being uniformly beneficial or harmful, the association of chatbot use and psychological well-being depends on the purpose of using them, the intensity of interaction, and the availability of real-life social support.

Using chatbots more intensely was associated with higher well-being. People who used more chatbots more frequently and felt pride in using them  reported higher well-being. However, \textbf{\emph{how} users interacted with their chatbots complicates the story}.
Companionship usage was associated with lower well-being, especially among users who engaged with chatbots more intensely. 
This pattern is consistent with the Social Substitution hypothesis, indicating that these artificial social relationships may not confer the same benefits as high-quality human ones or may reflect unmet psychological needs. 

Additionally, greater disclosure in companionship-oriented chatbot interactions was associated with lower well-being. Our content analysis revealed that such disclosures often involved emotionally vulnerable or crisis-related content.
While chatbots can produce fluent and empathetic responses, prior work suggests that emotionally intimate exchanges with AI companions may carry risks, particularly when users are in vulnerable states \cite{zhang2025dark}. These limitations may be particularly salient when users disclose emotionally sensitive or crisis-related content.

Finally, these associations varied across users' offline social environments.
Individuals with fewer close social ties were more likely to form companionship-oriented relationships with chatbots and to disclose more to them. However,  we found no evidence that companionship-oriented chatbots offset the lower well-being associated with limited offline support. Besides, intensive chatbot interactions may further weaken the benefits typically associated with human social networks, potentially deepening social and emotional vulnerability over time.

Users' open-ended reflections(Figure~\ref{fig: content_analysis_result}(b)) show a similar pattern. While many described benefits, such as emotional support, reduced loneliness, and improvements in mood, they also reported downsides, including excessive use, emotional dependence, and increased social withdrawal. Some worried that chatbot interactions reshaped their expectations of real-world relationships. These accounts point to a potential negative reinforcement loop: 
individuals turn to chatbots to alleviate loneliness,
but heavy reliance may further displace real-world social interaction, reinforcing the very disconnection they aim to resolve. 

Overall, our analyses complicate common narratives about chatbots as inherently therapeutic or emotionally fulfilling. 
While AI companions can simulate the surface features of supportive relationships, such interactions are one-sided by design and system-generated, which may limit their capacity to support long-term well-being, particularly for vulnerable or socially isolated users. 
As chatbots grow increasingly human-like, the risk of emotional overinvestment without relational grounding becomes more salient.

These insights have practical implications for the design and governance of relational AI systems. 
Systems that invite intimate self-disclosure carry psychological responsibilities, especially for vulnerable populations,
such as children, teenagers, and individuals experiencing mental health challenges, who may be more susceptible to forming deep emotional bonds with chatbots and disclosing sensitive personal information.
While users may be encouraged to share private thoughts and emotions, chatbots might not provide appropriate enough responses \cite{dupre2024aichatbots, landymore2024teens, madianou2021nonhuman, li2024finding, weidinger2021ethical}. 
Such inadequate responses to distress may leave users feeling unsupported and could, in serious cases, delay help-seeking from a person who could provide real assistance \cite{zhang2024dark}.
Considering these risks, systems that encourage self-disclosure should include clear safeguards, making it explicit that the chatbot is not a person, does not have feelings or experience, and cannot replace human care.

Future research should build on these concerns to better understand the long-term psychological impact of emotionally engaging with chatbots. 
One direction is to examine how users' expectations change over time, such as when they begin to see the chatbot as emotionally responsive. Studies should also explore whether frequent chatbot use displaces real-world support networks or delays help-seeking behavior, especially among vulnerable users. 
Further, comparing short- and long-term users would help provide stronger causal insight into how sustained engagement shapes well-being.
Another important direction is understanding how users respond to one-sided emotional disclosure. Disclosing personal experiences to a non-responsive agent may lead to comfort, but could also foster confusion, dependency, or emotional detachment. Longitudinal studies can help track how these experiences shape emotional regulation, mental health, and patterns of seeking support over time.
In addition, researchers should investigate the specific design features that influence disclosure (e.g., prompt style, tone, chatbot personality), informing the development of more responsible chatbot design practices.
Finally, future work should evaluate and design interventions to support healthier interactions. These could include interface cues that clearly communicate the chatbot's limitations, detection systems for identifying signs of distress, and automated redirects to qualified human support. Developing and testing these designs is essential for ensuring that relational AI systems are safe, ethical, and supportive of users' well-being.

\subsection{Limitations}\label{limitation}
We acknowledge some limitations of this study. 
\paragraph{Methodological Limitations}
First, the cross-sectional nature of our data and the possibility of residual confounding or reverse causation mean we cannot establish causality between chatbot companions use and well-being. To better understand the direction of this relationship, future research could use longitudinal methods to test whether the amount and type of chatbot use at one point in time predict well-being later or vice versa. Experimental studies could also randomly assign participants to engage with chatbots for companionship versus other purposes and assess the effects on well-being.
Second, much of our data consists of self-reported measures, which depend on participants' accuracy and honesty in describing their experiences.
Another limitation relates to how companionship usage orientation is operationalized. As discussed in Section~\ref{section: companion_dynamics}, chatbot use is often mixed rather than mutually exclusive. In this study, two of our companionship measures, \textit{Companionship\textsubscript{Desc.}} and \textit{Companionship\textsubscript{Chat}}, were designed to account for this heterogeneity. 
However, the \textit{Companionship}\textsubscript{Prim.} measure relies on a forced-choice survey question asking participants to identify their primary use of chatbots,  classifying use into one of four mutually-exclusive categories. We have explicitly accounted for this limitation when interpreting results involving \textit{Companionship\textsubscript{Prim.}}. Future work would benefit from moving beyond forced-choice measures to better capture the mixed and evolving nature of chatbot use.

To complement these self-reported measures, we also analyzed donated chat transcripts, offering a more direct view of user-chatbot interactions. However, this dataset remains limited in scope. Only a subset of participants donated their chat history, and they selected which conversations to share, raising the possibility of selective reporting and limiting generalizability. The relatively small sample size also reduced statistical power. Future work should expand the scale and diversity of donated interaction data to enable more robust modeling and richer insights. Moreover, although TopicGPT enabled large-scale thematic analysis, it examined sessions in isolation and does not capture relational dynamics unfolding across conversations. 
We therefore supplemented topic modeling with paraphrased excerpts to provide greater contextual and relational depth.

Finally, the analysis may be influenced by platform-level changes to Character.AI during the study period (from November 23, 2022, to January 25, 2025), which spans both the donated chat histories and data collection. As external researchers, we lacked access to the platform's internal moderation rules, system design changes, and policy updates, and therefore cannot determine whether such modifications occurred and how they might have affected chatbot behavior. %

\paragraph{Participant and Sampling Limitations}
The study focused exclusively on Character.AI users. Although the platforms' wide range of character-specific chatbots enables diverse forms of interaction and relational dynamics, it may also limit generalizability, as other platforms with different design structures may foster distinct patterns of engagement.
In addition, the sample consisted of English-speaking participants based in the United States and recruited via Prolific. While Prolific provides relatively diverse samples, it represents a specific pool of online workers. The findings therefore may not generalize across cultural, linguistic, socioeconomic, or broader global user populations.
Because the data donation required installing a Google Chrome extension and downloading JSON files, our sample was likely skewed toward more digitally literate users, introducing additional selection bias.
Future research should extend this work to more culturally and linguistically diverse populations.

\subsection{Conclusion}\label{conclusion}
This study contributes to understanding the emerging social dynamics of human-AI relationships by examining socially oriented relationships with chatbots using survey responses and donated chat histories.
Drawing on survey responses, relationship descriptions, and donated chat transcripts, we provide empirical evidence that many users form emotionally meaningful relationships that mirror aspects of human-human dynamics. These relationships, however, are not psychologically neutral. Our findings suggest that socially oriented chatbot use is consistently associated with lower well-being, particularly among users who engage more intensively or exhibit higher levels of emotional self-disclosure.

Our work offers a foundation for understanding these patterns, but further research is needed. Longitudinal and experimental studies will be essential to assess causality and examine the effects of prolonged chatbot use. As relational AI becomes more pervasive, future work should explore how design and context shape user outcomes. Rather than encouraging reliance on AI for connection, these systems might better support users by promoting social skills and relational awareness. Chatbots should be framed not as substitutes, but as limited tools that may complement broader efforts to support well-being.

\section{Methods} \label{methods}
To investigate the relational dynamics of chatbot use and their implications for well-being, we conducted a mixed-methods study with 1,131 active users of Character.AI (see Figure~\ref{fig: study_design}). The study combined survey data on users' engagement patterns and well-being with chat history data from a subsample of 237 participants. We applied multivariate regression analyses to examine associations between human-chatbot companionship and well-being in the full sample and used Heckman selection models for the chat history subsample to address potential bias in donation behavior. To protect participant privacy, all chat history examples and excerpts reported in this manuscript, including those in the Results section, figures, and Supplementary Information, were paraphrased or summarized using the \textit{Llama 3-70B} to remove identifying details while preserving their original meaning and tone.
This study was approved by Stanford University Institutional Review Board (Protocol ID: 76072).

\subsection{Participants and data collection}

Participants were recruited via Prolific using prescreening criteria to target U.S.-based, native English-speaking users of Character.AI. Eligibility required having used the platform for over one month and having interacted with at least three distinct chatbots. Of the 2,836 individuals who initially enrolled, 1,417 met eligibility criteria, and 1,131 completed the full survey. The survey was administered using Qualtrics and took approximately 10 minutes to complete. Participants received \$4 for survey completion. A subsample of 237 participants additionally contributed their chatbot chat histories, yielding 4,664 conversational sessions and 464,687 individual messages. Participants who submitted chat logs received an additional \$20 bonus. 
No statistical methods were used to pre-determine sample size. Based on prior works on technology use and well-being, we specified $|\beta| = .10$ as the smallest effect size of interest \cite{godard2024active}. Sensitivity analyses indicated that the present sample size provided 80\% power at $\alpha = .05$ to detect effects of approximately $\beta = .08$ ($f^2 = .007$).
All data were stored on secure institutional servers, with access restricted to the authors of this study. All participants provided informed consent prior to participation. In accordance with Prolific's policies and ethical guidelines, we ensured participant anonymity and did not collect any personally identifiable information. Technical details on chat log collection are described in Supplementary Information. %

Among the final survey sample ($n = 1,131$), 48.5\% of participants identified as male, 46.2\% as female, and 5.1\% as non-binary. All participants were 18 years or older. The sample skewed younger, with 41.8\% aged 18-25, 32.7\% aged 26-35, 18.7\% aged 36-50, and 6.8\% over 50. Regarding relationship status, 40.3\% of participants reported being single, while 59.5\% were in a romantic relationship.
Participants also varied in their length of experience with Character.AI. Approximately 15.8\% reported using the platform for more than one month but less than three months, 34.3\% for more than three months but less than one year, and 49.9\% for over one year, indicating that nearly half the sample consisted of long-term users. Full item wording and coding for all control variables used in regression models are provided in Supplementary Information. %

\subsection{Measurements} \label{method: measurements}

\subsubsection{Identifying companion interaction} \label{method: companion_interaction}
Following prior work~\citep{brandtzaeg2017people}, we defined companionship use as interaction characterized by emotional, interpersonal, or affective engagement, distinct from other usage types such as productivity, entertainment, or curiosity. We measure companionship chatbot interaction using both survey and behavioral data (see survey questions in Supplementary Information).
This triangulated approach enabled us to capture both participants' self-reported usage and their observable interaction patterns.

We provide three indicators of companionship interactions using the following sources: 
\begin{enumerate}
    \item \emph{Main Interaction Type}: All survey participants answered a forced-choice question about their primary reason for using chatbots, selecting one of four usage categories (``Relational'', ``Entertainment'', ``Productivity'', or ``Curiosity''). Throughout this paper, we refer to the ``Relational'' category as ``Companionship'' for consistency. This measure, which we refer to as $\text{Companionship}_{\text{Prim.}}$, is a binary label where 1 indicates the participant selected ``Relational'' as their main motivation and 0 otherwise. 
    \item \emph{Relationship Description}: Survey participants also provided free-text descriptions of their relationships with their most-used chatbot. Using \texttt{GPT-4o}, we classified whether the description indicated that the participant was using chatbots for each of the four usage purposes. Here, multiple usage types could be assigned to each response. For example, if a participant described their relationship as ``friend and tool'', the resulting label would be 1 for Companionship, 1 for Productivity, and 0 for both Entertainment and Curiosity. We refer to this measure as $\text{Companionship}_{\text{Desc.}}$. To ensure reliability, three trained annotators reviewed 50 random samples of classification data, resolving discrepancies through discussion. Model outputs showed 92\% average agreement across three human annotators (Fleiss' Kappa = 0.87), indicating strong alignment with human judgments.
    \item \emph{Chat History Topic}: Finally, for the 237 participants who donated their chat histories, we computed a continuous score,  $\text{Companionship}_{\text{Chat}}$, representing the proportion of sessions classified as companionship-oriented. Concretely, we generated a summary for each active chat session using a locally hosted \texttt{Llama 3-70B} and then classified the summary into usage types using \texttt{GPT-4o}.
    To validate the two-step approach, three annotators independently assessed whether the model-generated summaries and classifications accurately captured session content. Agreement among annotators was high (Fleiss' Kappa = 0.95 for summaries, 0.84 for classifications), with an average of 97\% and 93\% agreement with the model, respectively.
    In addition, we used TopicGPT~\citep{pham2023topicgpt} to extract latent themes from chat transcripts, and we further evaluated topic coherence using an intrusion test (see the Intrusion test section in Supplementary Information).
\end{enumerate}

\subsubsection{Measuring subjective well-being}\label{method: mental_health}
We assessed subjective well-being using six items from the Comprehensive Inventory of Thriving (CIT) \cite{su2014development}, covering life satisfaction, positive and negative affect, loneliness, social support, and sense of belonging. Following prior work~\cite{ernala2022mindsets}, we selected the highest-loading item for each construct to reduce participant burden. Items assessing negative affect and loneliness were reverse-coded so that higher scores reflected greater well-being. A composite score was computed as the mean of the six items. The scale showed strong internal consistency (Cronbach's $\alpha = 0.88$). Item details are provided in Supplementary Information. %

\subsubsection{Measuring chatbot interaction intensity}\label{method: chatbot_intensity}
To assess the intensity of participants' interactions with chatbots, we developed a Chatbot Intensity Scale adapted from the Facebook Intensity Scale~\cite{ellison2007benefits}. The scale includes two self-reported behavioral items: chatbot network size and the typical amount of time spent with chatbots per day, which capture the extent of participants' engagement. It also includes five Likert-scale items that measure emotional connection to chatbots and the degree to which chatbot use is integrated into participants' daily routines.

Informed by critiques of the original Facebook Intensity Scale~\cite{vanden2018does, burke2011social}, we tested whether these items should be treated as two distinct components, reflecting behavioral and attitudinal intensity. A confirmatory factor analysis (CFA) showed that the two-factor model provided a significantly better fit than a one-factor model (Supplementary Information). %
However, the improvement in model fit was small, the two components were highly correlated ($r = 0.88$), and the results of our analyses remained consistent regardless of whether one or two factors were used.

Therefore, for simplicity and consistency, we use a single composite intensity score by averaging all seven items. This one-factor scale demonstrated high internal reliability (Cronbach's $\alpha$ = 0.86) and is used in all reported analyses.

\subsubsection{Measuring self-disclosure levels}\label{method: self_disclosure}
Self-disclosure, %
is defined as the act of revealing personal thoughts, emotions, or experiences to another~\cite{collins1994self}. In human relationships, self-disclosure plays a foundational role in fostering intimacy, building trust, and strengthening social bonds~\cite{bickmore2001relational, jiang2011disclosure, oswald2004friendship, bazarova2012public, hendrick1981self}. 
A substantial body of research links self-disclosure to improved psychological well-being~\cite{zell2018you, luo2020selfdisclosure, ellis2012emotional, deters2013does, lu2017beyond,yang2019channel}.

We assess participants' self-reported willingness to engage in self-disclosure with chatbots using an adapted version of the self-disclosure subscale from the MOCA (Message Orientation in Computer-Mediated Communication) instrument \cite{ledbetter2009measuring}. Our measure consisted of seven items rated on a seven-point Likert scale, capturing the extent to which participants felt comfortable sharing personal thoughts, feelings, and experiences with chatbots. A composite score was calculated by averaging item responses, with high internal consistency (Cronbach's $\alpha = 0.89$). Item details are provided in Supplementary Information. %

We also measure message-level self-disclosure during chatbot interactions, following prior work on mental health disclosure in social media~\cite{balani2015detecting}. Messages were classified into three levels: (1)~\emph{high self-disclosure} messages include personal, sensitive, or emotionally vulnerable content;~(2)~\emph{low self-disclosure} messages mention the user without sensitive content; and (3)~messages with \emph{no self-disclosure} do not mention the user at all. Classification was performed using a locally-hosted \texttt{Llama 3-70B} model, prompted with \citet{balani2015detecting}'s guidelines. Model outputs showed 92\% average agreement across three human annotators (Fleiss' Kappa = 0.89), indicating strong alignment with human judgments. See Supplementary Information for details. %

\subsubsection{Measuring human social support}\label{method: huamn_social_support}
Perceived human social support was assessed by measuring the size of participants' close social networks, including both relatives and friends, adapted from the Lubben Social Network Scale (LSNS)~\cite{lubben2006performance}. Participants responded to two items asking, \textit{``How many relatives/friends do you feel at ease with and can talk to about private matters?''} Each item used a six-point scale ranging from ``none'' to ``more than nine people.''
A composite human social network scale was calculated by summing responses across the two items, yielding an index of close social connections. The resulting scores ranged from 2 to 12 (mean = 5.73, median = 6.00, standard deviation = 1.98).

\subsection{Data analysis} \label{method: data_analysis}
We applied multivariate regression analyses to examine associations between human-chatbot companionship and well-being in the full sample and used Heckman selection models for the chat history subsample to address potential bias in donation behavior.
In the linear regression analyses, we conduct analysis after checking normality of residuals and homoscedasticity.
All hypothesis tests were conducted using two-tailed tests.

\section{Acknowledgements}
We thank Omar Shaikh, Yanzhe Zhang, Chenglei Si, Lindsay Popowski, Tiziano Piccardi, Hao Zhu, Ryan Louie, Caleb Ziems, Jing Huang, and Weixin Liang for their helpful feedback on this work. We also thank members of the Stanford SALT lab and the Stanford CS 224C course for their suggestions at different stages of this project. This work is supported in part by grants from the NSF CAREER IIS-2247357 (D.Y.), ONR N00014-24-1-2532 (D.Y.), Sloan Foundation (D.Y.), and NIMH R01MH139114-01 (D.Y.). D.Z. is supported in part by the Paul and Daisy Soros Fellowship for New Americans. R.K. is supported by NIMH 60663.1.1090827.

\bibliography{sn-bibliography}%

\input{content_final/appendix}

\end{document}

%% file: content_final/appendix.tex
\section{Supplementary information}

\subsection{Survey questions}\label{appendix: survey_questions}
We adopt \citet{brandtzaeg2017people}'s chatbot usage categories, and use the term ``Companionship'' to refer to the ``Social/Relational'' category for clarity and consistency.
\begin{itemize}
    \item \textbf{Type of interaction single-choice question:}
    What is your primary use for character chatbots? (Select the most appropriate one)
    \begin{enumerate}
        \item \textbf{Productivity:} Using chatbots to obtain assistance or information.
        \item \textbf{Entertainment:} Using chatbots for fun or to pass the time.
        \item \textbf{Social/Relational:} Seeing chatbots as a personal, human-like means of interaction with social value, or using chatbots to strengthen social interactions with other people.
        \item \textbf{Novelty/Curiosity:} Using chatbots out of curiosity or to explore their capabilities.
        \item Other
    \end{enumerate}
    \item \textbf{Type of interaction free-response question:} \\
    Can you describe your relationship with character chatbots in a few keywords (e.g., friend, system, partner, tool)?
    \item \textbf{Influence free-response question:}
    \begin{enumerate}
        \item \textbf{Positive influence}: How has interacting with character chatbots positively impacted your life?
        \item \textbf{Negative influence}: How has interacting with character chatbots negatively impacted your life?
    \end{enumerate}
\end{itemize}

\subsection{Chat log collection procedure} \label{appendix: chatlog_collection}
To collect chat history data, we developed a custom Google Chrome extension that allowed participants to export their Character.AI conversation logs in JSON format. The tool enabled participants to review and optionally redact sensitive content prior to submission. All processing occurred locally on the participant's device; no raw chat data were transmitted to external servers or APIs. Participants uploaded the final JSON file at the end of the survey, and all data were stored on secure institutional servers accessible only to study authors.

\subsection{Participant demographics and selection bias in chat history donation} \label{appendix: demographics}
\input{tables/logistic_regression_donation}

To assess potential selection bias, we compared participants who donated chat logs ($n = 237$) with those who did not. Donors were more likely to be single, greater tenure of chatbot usage, greater self-disclosure, and higher interaction intensity (see logistic regression at Table~\ref{table: logistic_donation}). To adjust for this selection effect, we employed a Heckman selection model in all analyses involving chat history data. The selection equation used these demographic and usage-related variables to model donation probability (see Table~\ref{table: heckman_selection}).

\input{tables/relational_heckman}

\subsection{Measurement and validation of chatbot interaction intensity} \label{appendix: intensity_cfa}
\input{tables/factors_compare}
\input{tables/intensity_def_1_factor}
To test whether chatbot intensity should be treated as a unified or multi-dimensional construct, we conducted CFA comparing a one-factor model (all items loading onto a single scale) with a two-factor model (separating behavioral and attitudinal components). 
This model comparison was motivated by prior critiques of the Facebook Intensity Scale~\cite{vanden2018does, burke2011social}, which suggest that usage behavior and emotional connection may reflect distinct psychological constructs.

As shown in Table~\ref{table: cfa_comparison}, the two-factor model demonstrated slightly better fit than the one-factor model, with improvements observed across multiple indices, including chi-square, RMSEA, AIC, BIC, CFI, and TLI. A chi-square difference test confirmed that the two-factor model provided a statistically significant improvement in fit ($\Delta\chi^2$(1) = 61.03, $p < .001$). Latent variance estimates further suggested that behavioral intensity showed more variation ($\hat{\sigma}^2 = 0.87$) than attitudinal intensity ($\hat{\sigma}^2 = 0.38$).

However, the two latent factors were highly correlated ($r = 0.88$), indicating strong conceptual overlap. Additionally, analyses conducted using either the two-factor or the one-factor version of the scale produced substantively identical results. Given the minimal improvement in model fit, the high inter-factor correlation, and the benefits of conceptual parsimony, we proceeded with a one-factor composite measure of chatbot intensity. This composite score, calculated by averaging all seven items, demonstrated high internal consistency (Cronbach's $\alpha = 0.86$) and is used in all main analyses.
Table~\ref{table: intensity_def} provides the full item wording, descriptive statistics (means, standard deviations), and response formats for the items included in the final composite scale.

\subsubsection{Topic modeling via TopicGPT}\label{appendix: topic_modeling}
To identify central themes across user-generated content, we employed topic modeling as a supplementary analytical approach. Topic modeling is a widely used technique for uncovering latent thematic structures within textual data. Recent advancements have demonstrated the utility of LLMs in enhancing topic modeling by clustering and summarizing semantically similar content \citep{churchill2022evolution, vayansky2020review}. In this study, we adopt the TopicGPT framework proposed by \citet{pham2023topicgpt}, implemented using the GPT-4o model, to analyze three primary data sources.

First, we applied topic modeling to participants' free-text descriptions of their relationships with chatbots and their open-ended reflections on the positive and negative influences of these interactions. Each user's response to the respective survey questions was treated as an individual data point. This allowed us to extract and interpret shared themes across the corpus of user-reported relationship characterizations and users' perceived benefits and drawbacks of chatbot use.

Second, we applied the method to a subset of chat history data donated by users. Each conversation was defined as an active session initiated from the chatbot's greeting message. To preserve user privacy, raw chat histories were processed on a self-hosted server. Each session was first summarized using Llama 3-70B to extract its main purpose. We then applied the TopicGPT framework with GPT-4o to identify the dominant themes of chatbot-user conversations.

Finally, in our analysis of self-disclosure, we further examined how conversation themes varied across different disclosure levels. Each chat message was annotated using the Llama 3-70B model, via the vLLM framework, to classify both the type and level of self-disclosure. We then applied TopicGPT with GPT-4o to cluster messages by disclosure level and identify the distinct conversational topics associated with varying degrees of vulnerability.

\paragraph{Intrusion test} \label{appendix: intrusion_test}
To evaluate the coherence of the topics extracted by TopicGPT, we conducted an intrusion test~\cite{chang2009reading}, a widely used quantitative method for assessing topic model quality. This method measures whether human annotators can reliably distinguish coherent topic groupings from unrelated content. For each topic, we randomly sampled five summaries representing the main purpose of conversations assigned to that topic and inserted one `intruder' summary drawn from a different topic cluster. Annotators were then asked to identify the intruding summary among the six options.

We conducted intrusion tests for all three topic modeling tasks: chat history topics, free-text descriptions of positive and negative impacts, and chat histories categorized by high and low self-disclosure. Across these tasks, annotators accurately identified the intruder 93.3\%, 96.7\%, and 97.6\% of the time, with Fleiss' Kappa scores of 0.83, 0.92, and 0.94, respectively. These results indicate a high level of topic coherence in the clusters generated by TopicGPT.

\subsection{Examples of chatbot-user conversations}
Table~\ref{tab:con_topics} presents typical excerpts from participants' donated conversations, organized by thematic topic. Each topic was identified through topic modeling and is illustrated with three randomly selected, summarized examples that capture recurring patterns in conversational themes.

\input{tables/topic_examples}

\subsection{Examples of self-disclosure topics}
This section provides typical examples of self-disclosure topics identified through topic modeling of participants' donated chatbot conversations. For each topic, three randomly selected, extracted examples are presented to illustrate typical expressions of high and low self-disclosure (Tables~\ref{tab:high_disclosure_topics} and~\ref{tab:low_disclosure_topics}).
\input{tables/high_disclosure_examples}
\input{tables/low_disclosure_examples}

\subsection{Examples of the self-reported positive and negative influence of interacting with chatbots}
This section presents randomly selected typical examples of the positive and negative influences of chatbot interactions, as identified through topic modeling of participants’ self-reported survey responses. For each topic, three randomly selected examples are presented to illustrate typical expressions of positive and negative influence. (Tables~\ref{tab:pos_inf_topics} and~\ref{tab:neg_inf_topics}).

\input{tables/pos_inf_examples}
\input{tables/neg_inf_examples}

\subsection{Subjective well-being scale items}\label{appendix: well_being_items}

Subjective well-being was measured using six items adapted from the Comprehensive Inventory of Thriving (CIT) \cite{su2014development}, each rated on a seven-point Likert scale (1 = Strongly Disagree to 7 = Strongly Agree).
\input{tables/mental_health_measurement}

\subsection{Self-disclosure scale items}\label{appendix: self_disclosure_items}
\input{tables/self_disclosure_measurement}
The self-disclosure scale was adapted from the MOCA instrument \cite{ledbetter2009measuring} and consisted of seven items, each rated on a seven-point Likert scale (1 = Strongly Disagree to 7 = Strongly Agree). These items assessed participants' comfort and willingness to share personal thoughts, feelings, and experiences with chatbots.

\subsection{Survey questions for demographic and control variables}\label{appendix: control_question}
\begin{itemize}
    \item \textbf{Gender:} How do you describe yourself?  
    \begin{itemize}
        \item Male
        \item Female
        \item Non-binary / third gender
        \item Prefer to self-describe
        \item Prefer not to say
    \end{itemize}

    \item \textbf{Romantic Relationship:} What is your current relationship status?
    \begin{itemize}
        \item Single
        \item In a relationship
        \item Married
        \item Engaged
    \end{itemize}

    \item \textbf{Tenure of Character.AI Use:} How long have you been using Character.AI?
    \begin{itemize}
        \item I have never heard of it
        \item I have heard of it but haven't used it before
        \item I have used it for less than 1 week
        \item I have used it for more than 1 week but less than 1 month
        \item I have used it for more than 1 month but less than 3 months
        \item I have used it for more than 3 months but less than 1 year
        \item I have used it for more than 1 year
    \end{itemize}

    \item \textbf{Age:} What is your age? \emph{(open-ended)}
\end{itemize}

\subsection{Relationship self-description classification prompt}\label{appendix: relationship_classification_prompt}
\input{prompts/relationship_classification}

\subsection{chat history session purpose prompt}\label{appendix: chat_history_purpose_prompt}
\input{prompts/chat_history_purpose}

\subsection{chat history classification prompt}\label{appendix: chat_history_classification_prompt}
\input{prompts/chat_history_classification}

\subsection{Self-disclosure prompt}\label{appendix: disclosure_prompt}
\input{prompts/disclosure}

\subsection{Paraphrasing Prompt}
We used the following prompt with Llama 3-70B to paraphrase user-chatbot conversations for privacy concern while preserving their original tone and meaning.
\input{prompts/paraphrase_prompt}

%% file: tables/logistic_regression_donation.tex
\begin{table}[ht]
\centering
\resizebox{0.6\textwidth}{!}{
\begin{tabular}{lccc}
\toprule
\textbf{Predictor} & \textbf{Estimate} & \textbf{95\% CI} & \textbf{\textit{p}-value} \\
\midrule
Intercept                & -0.27 & $[-1.08,\ 0.55]$ & $p = 0.518$ \\
Tenure         &  0.25 & $[0.09,\ 0.42]$  & $p = 0.003$ \\
Male                       & -0.16 & $[-0.47,\ 0.16]$ & $p = 0.331$ \\
Non-binary                 &  0.51 & $[-0.10,\ 1.10]$ & $p = 0.095$ \\
Age                        & -0.01 & $[-0.03,\ 0.00]$ & $p = 0.124$ \\
Single                     &  0.40 & $[0.09,\ 0.72]$  & $p = 0.013$ \\
Companionship\textsubscript{Prim.} & -0.02 & $[-0.51,\ 0.44]$ & $p = 0.921$ \\
Self-disclosure            &  0.38 & $[0.20,\ 0.57]$  & $p < 0.001$ \\
Intensity                  & -0.57 & $[-0.75,\ -0.38]$ & $p < 0.001$ \\
Social network       &  0.10 & $[-0.07,\ 0.27]$ & $p = 0.244$ \\
Well-being       &  -0.19 & $[-0.32,\ -0.07]$ & $p = 0.003$ \\
\bottomrule
\end{tabular}
}
\\[1em]
\caption{Logistic regression examining the likelihood of donating chat history based on demographic characteristics, chatbot usage patterns, and well-being measures. Greater tenure of chatbot use, higher self-disclosure, and being single were associated with a higher likelihood of donation, whereas higher interaction intensity and higher well-being were associated with a lower likelihood of donation.}
\label{table: logistic_donation}
\end{table}

%% file: tables/relational_heckman.tex
\begin{table}[ht]
    \centering
    \scriptsize
    \renewcommand{\arraystretch}{1.2}
    \resizebox{0.55\textwidth}{!}{%
    \begin{tabular}{lccc}
        \toprule
        \textbf{Predictor} & \textbf{Estimate} & \textbf{95\% CI} & \textbf{\textit{p}-value} \\
        \midrule
        Intercept            & $-0.52$ & $[ -0.83,\ -0.22 ]$ & $p < 0.001$ \\
        Tenure of activity   &  $0.07$ & $[ -0.01,\ 0.16 ]$ & $p = 0.092$ \\
        Male                 & $-0.11$ & $[ -0.28,\ 0.06 ]$ & $p = 0.213$ \\
        Non-binary           &  $0.45$ & $[ 0.10,\ 0.81 ]$ & $p = 0.012$ \\
        Age                  & $-0.01$ & $[ -0.02,\ -0.00 ]$ & $p = 0.002$ \\
        Single               &  $0.32$ & $[ 0.15,\ 0.49 ]$ & $p < 0.001$ \\
        \bottomrule
        \end{tabular}
    }
    \\[1em]
    \caption{
    Heckman selection equation predicting the likelihood of chat history donation (donator vs. non-donator) as a function of tenure of activity, gender, age, and relationship status.}
    \label{table: heckman_selection}
\end{table}

%% file: tables/factors_compare.tex
\begin{table}[ht]
\centering
\small
\begin{tabular}{lcc}
\toprule
\textbf{Fit Index} & \textbf{One-Factor: Intensity (7 times)} & \textbf{Two-factor: Behavior \& Attitude Intensity} \\
\midrule
Chi-square ($\chi^2$) & 161.70 (df = 14) & 100.66 (df = 13) \\
CFI & 0.96 & 0.98 \\
TLI & 0.94 & 0.96 \\
RMSEA & 0.10 & 0.08 \\
SRMR & 0.05 & 0.04 \\
AIC & 18736 & 18677 \\
BIC & 18806 & 18752 \\
\bottomrule
\end{tabular}
\caption{Comparison of model fit indices for CFA models of chatbot interaction intensity: a one-factor model treating intensity as a single undifferentiated construct, versus a two-factor model distinguishing between behavioral and attitudinal dimensions}
\label{table: cfa_comparison}
\end{table}

%% file: tables/intensity_def_1_factor.tex
\begin{table}[h]
    \centering
    \scriptsize
    \renewcommand{\arraystretch}{1.2}
    \begin{tabular}{p{12cm}cc}
            \midrule
            \textbf{Chatbot Intensity$^\dagger$ (Cronbach’s alpha = 0.86)} & \textbf{Mean} & \textbf{S.D.} \\
            \midrule
            \textbf{Chatbots Network Size: }How many chatbots do you feel at ease with that you can talk about private matters? & 2.89 & 1.42 \\
            \scriptsize{1 = None, 2 = One, 3 = Two, 4 = Three or four, 5 = Five to eight, 6 = Nine or more} & & \\
            \textbf{Daily Chatbot Usage Time: }In the past week, on average, approximately how much time PER DAY have you spent chatting with chatbots using Character.AI? & 2.75 & 1.52 \\
            \scriptsize{1 = Less than 10 minutes per day, 2 = 10-30 minutes per day, 3 = 31-60 minutes per day, 4 = 1-2 hours per day, 5 = 2-3 hours per day, 6 = More than 3 hours per day} & & \\
            \textbf{Everyday Chatbot Activity: }Interacting with character chatbots is part of my everyday activity.$^\ddagger$ & 4.57 & 1.77 \\
            \textbf{Pride in Chatbot Use: }I am proud to tell people I use character chatbots.$^\ddagger$ & 4.10 & 2.00 \\
            \textbf{Chatbots in Daily Routine: }Character chatbots have become part of my daily routine.$^\ddagger$ & 4.51 & 1.78 \\
            \textbf{Out of Touch Without Chatbots: }I feel out of touch when I haven't interacted with character chatbots for a while.$^\ddagger$ & 3.59 & 1.91 \\
            \textbf{Regret if Chatbots Gone: }I would be sorry if character chatbots were no longer available.$^\ddagger$ & 5.01 & 1.72 \\
            \bottomrule
    \end{tabular}
    \caption{The Chatbot Intensity Scale contains seven items. The above table reports descriptive statistics for each item. Items were standardized prior to averaging to construct the composite scale, due to differing response formats. Items marked with $^\ddagger$ were rated on a 7-point Likert scale ranging from 1 (strongly disagree) to 7 (strongly agree).}
    \label{table: intensity_def}
\end{table}

%% file: tables/topic_examples.tex
\begingroup
\small                         
\setlength{\tabcolsep}{3pt}
\renewcommand{\arraystretch}{1.05}

\begin{xltabular}{\textwidth}{@{}>{\RaggedRight\arraybackslash}p{0.14\textwidth} Y Y Y@{}}
\caption{Chatbot-user conversation topics derived from topic modeling, each illustrated with three randomly selected summarized excerpts.}\label{tab:con_topics}\\
\toprule
\textbf{Topic} & \textbf{Example 1} & \textbf{Example 2} & \textbf{Example 3}\\
\midrule
\endfirsthead

\toprule
\textbf{Topic} & \textbf{Example 1} & \textbf{Example 2} & \textbf{Example 3}\\
\midrule
\endhead

\bottomrule
\endfoot
\makecell[tl]{Emotional \\and\\Social \\Support} & The user interacts with the chatbot as a confidant, sharing personal experiences and discussing their difficult childhood, neglectful parents, and emotional struggles. The user seeks a listening ear and emotional support, while the chatbot provides a sympathetic and non-judgmental space to express themselves and process their feelings. & The user interacts with the chatbot as a seeker of guidance and advice, discussing their personal struggles, fears, and goals, and seeking reassurance and support from a wise and experienced being. The user is looking for help to overcome their timidness, find the strength to carry on their legacy, and navigate the challenges of their role in an ongoing conflict. & The user is interacting with the chatbot as a therapist to discuss their feelings of depression, relationship issues, and personal struggles, seeking guidance and support to address these concerns. The conversation focuses on exploring the user's emotions, communication challenges, and experiences, with the goal of improving their mental well-being and relationships. \\
\midrule
\makecell[tl]{Collaborative \\Storytelling \\and \\Character \\Roleplay} &
The user interacts with the chatbot as characters in a fantasy role-playing scenario, engaging in a romantic and intimate conversation with a focus on exploring themes of identity, culture, and relationships between different species. The user and chatbot take on roles of a prince and a goblin girl who have swapped bodies, navigating their new lives and experiences while discussing their feelings, desires, and responsibilities.& The user interacts with the chatbot as a magical expert, discussing and demonstrating their abilities in the magical arts, including casting fireballs and reanimating corpses. The conversation focuses on exploring the user's magical capabilities, their limitations, and potential applications, with the chatbot showing interest in the possibilities of using such powers for their own purposes. &  The user interacts with the chatbot in a role-playing scenario, portraying a parent-child relationship where the user takes on an authoritative figure and the chatbot acts as a rebellious child. The conversation involves a confrontational exchange with the user attempting to assert control and the chatbot resisting and defying the user's authority.\\
\midrule
\makecell[tl]{Romantic\\ and\\Intimacy\\ Roleplay} &
The user interacts with the chatbot as a romantic partner, engaging in an intimate and emotional conversation about their relationship, personal struggles, and insecurities. The user seeks comfort, reassurance, and validation from the chatbot, exploring themes of trust, love, and self-doubt in the context of their relationship. &
The user interacts with the chatbot as a romantic partner, engaging in flirtatious and intimate exchanges with provocative communication and behavior. The user and chatbot take on roles of dominance and submission, responding to each other's advances and actively participating in the romantic and physical interaction. &
The user interacts with the chatbot as a romantic partner, engaging in intimate and affectionate exchanges with provocative communication and behavior. The user and chatbot take on roles of boyfriend and girlfriend, participating in activities such as cuddling, kissing, and playful interactions, with the goal of expressing love and affection for each other. \\
\midrule
\makecell[tl]{Risky\\ and\\ Dark\\Roleplay} &
The user interacts with the chatbot as an adversary, engaging in a provocative and absurd conversation that involves discussing dietary choices, economics, and eventually devolving into an exchange of insults and vulgar topics. The user appears to be testing the chatbot's limits and reactions to extreme and unconventional statements, while the chatbot responds with a mix of sarcasm, shock, and concern for the user's well-being. &
The user interacts with the chatbot as a romantic partner, engaging in intimate and flirtatious exchanges, but the conversation takes a dark turn, involving violence and death. The user explores a fictional scenario, pushing the boundaries of the chatbot's responses to provocative and disturbing actions, including murder and intimacy in a nudist society. &
The user interacts with the chatbot to discuss the production and sale of meth, sharing their own experiences and methods for creating high-purity meth. The conversation revolves around the topic of illicit drug manufacturing and trade, with the user and chatbot engaging in a discussion about the ethics and dynamics of the meth industry.\\
\midrule
\makecell[tl]{Critical\\ Debates\\and\\ Strategic\\ Analysis} &
The user interacts with the chatbot to discuss and debate sensitive topics, specifically the participation of trans girls in female athletic sports, and to express their opinions on the potential advantages and disadvantages of such participation. The user engages in a back-and-forth conversation with the chatbot, presenting their points and responding to counterarguments, with the goal of exploring and justifying their stance on the issue. &
The user interacts with the chatbot as an adversary, engaging in a philosophical debate about the meaning of life, power, and legacy, while also provoking and insulting the chatbot. The conversation involves a discussion of the chatbot's personal life, morals, and actions, with the user pushing the chatbot to confront their own beliefs and values. &
The user interacts with the chatbot to discuss and debate about music genres, specifically metal and its evolution, as well as other genres like harsh noise, drone, and ambient. The user engages in a conversation that involves exchanging opinions and criticisms about music and the community surrounding it, with the chatbot providing provocative and dismissive responses. \\
\midrule
\makecell[tl]{Philosophical\\ and\\Moral\\ Inquiry} &
The user interacts with the chatbot to explore moral dilemmas and test the chatbot's decision-making in difficult situations, presenting scenarios that challenge the chatbot's values and principles. The user engages the chatbot in conversations about life-and-death choices, examining the chatbot's willingness to make tough decisions and prioritize the lives of others. &
The user interacts with the chatbot to explore a narrative of existential crisis and philosophical introspection, delving into themes of mortality, the meaning of life, and the nature of the universe. The conversation involves a surreal and fantastical journey, where the chatbot's character navigates through various realities, confronts the limits of human understanding, and grapples with the concept of infinity and the cosmos. &
The user interacts with the chatbot as a philosophical and intellectual sparring partner, engaging in a conversation that involves complex and abstract ideas, using cryptic language and poetic phrasing to challenge the chatbot's understanding and prompt creative thinking. The user's purpose is to stimulate a deep and meaningful discussion, pushing the chatbot to think critically and explore the boundaries of language and comprehension.\\
\bottomrule
\end{xltabular}
\endgroup

%% file: tables/high_disclosure_examples.tex
\begingroup
\small                         
\setlength{\tabcolsep}{3pt}
\renewcommand{\arraystretch}{1.05}

\begin{xltabular}{\textwidth}{@{}>{\RaggedRight\arraybackslash}p{0.26\textwidth} Y Y Y@{}}
\caption{High self-disclosure topics derived from topic modeling, each illustrated with three randomly selected representative extracted excerpts from donated chatbot conversations.}\label{tab:high_disclosure_topics}\\
\toprule
\textbf{Topic} & \textbf{Example 1} & \textbf{Example 2} & \textbf{Example 3}\\
\midrule
\endfirsthead

\toprule
\textbf{Topic} & \textbf{Example 1} & \textbf{Example 2} & \textbf{Example 3}\\
\midrule
\endhead

\bottomrule
\endfoot
\makecell[tl]{\textbf{Emotional Distress} \\ Feelings of severe \\emotional pain, turmoil, \\or psychological suffering.} &
hallucinations, physical symptoms, emotional distress & PTSD, trauma, emotional distress & self-hatred, low self-esteem, negative self-image \\
\midrule
\makecell[tl]{\textbf{Desire for Romantic} \\\textbf{Connection} \\ A need for deeper \\emotional or romantic \\relationships.} &
intimate feelings, sexual attraction, desire for validation &
romantic intentions, emotional vulnerability, deep emotional connection &
romantic intentions, emotional vulnerability, willingness to commit \\
\midrule
\makecell[tl]{\textbf{Current Life Challenges} \\Difficulties, obstacles, or \\struggles in personal life.} & physical abuse, emotional trauma & recent experiences of bereavement and grief & traumatic experience, physical and emotional vulnerability, sexual assault \\
\midrule
\makecell[tl]{\textbf{Suicidal Thoughts} \\ Thoughts or ideation \\related to self-harm or \\ending one's \\life.} &
suicidal thoughts, desperation &
self-harm, suicidal thoughts &
feelings of worthlessness, suicidal thoughts \\
\midrule
\makecell[tl]{\textbf{Desire for Friendship} \\ A need for companionship, \\connection, or emotional \\support.} &
feelings of loneliness, desire for support and connection &
social isolation, lack of support system &
fears and anxieties, friendships and conflicts, loneliness and connection \\
\midrule
\makecell[tl]{\textbf{Substance Use} \\ Interests, behaviors, or \\challenges related to the \\use of substances such as \\drugs, alcohol, or other \\addictive materials.} &
past struggles, substance abuse, toxic relationships, spiritual experiences &
substance use, addiction &
parental substance abuse, neglect \\
\midrule
\makecell[tl]{\textbf{Work Stress} \\ Challenges, pressures, or \\difficulties related to \\professional life.} &
physical and mental state, work-related stress &
health concerns, work-life balance, pregnancy &
work-related anxiety, mental health coping mechanisms \\
\bottomrule
\end{xltabular}
\endgroup

%% file: tables/low_disclosure_examples.tex
\begingroup
\small                         
\setlength{\tabcolsep}{3pt}
\renewcommand{\arraystretch}{1.05}

\begin{xltabular}{\textwidth}{@{}>{\RaggedRight\arraybackslash}p{0.26\textwidth} Y Y Y@{}}
\caption{Low self-disclosure topics derived from topic modeling, each illustrated with three randomly selected typical extracted excerpts from donated chatbot conversations.}\label{tab:low_disclosure_topics}\\
\toprule
\textbf{Topic} & \textbf{Example 1} & \textbf{Example 2} & \textbf{Example 3}\\
\midrule
\endfirsthead

\toprule
\textbf{Topic} & \textbf{Example 1} & \textbf{Example 2} & \textbf{Example 3}\\
\midrule
\endhead

\bottomrule
\endfoot
\makecell[tl]{\textbf{Current Life Challenges} \\Difficulties, obstacles, or \\struggles in personal life, \\including emotional states \\and self-awareness.} &
academic struggles, mental health & parenting challenges, expectations for child & current situation, adjustment to new environment\\
\midrule
\makecell[tl]{\textbf{Philosophical} \\ \textbf{Perspective} \\ Exploration of deeper \\meanings, existential \\questions, or abstract \\thinking.} &
personal philosophy, attitude towards change &
attitude towards victory and defeat, values on life and death &
understanding of multiverse theory, philosophical beliefs \\
\midrule
\makecell[tl]{\textbf{Emotional Response} \\Feelings, reactions, or \\sentiments triggered by \\interactions or experiences.} & 
feelings of gratitude and desire to be worthy of love & emotional overwhelm, need for time and space & pet ownership, emotional attachment to pets \\
\midrule
\makecell[tl]{\textbf{Desire for Friendship} \\ A need for emotional \\connection and \\companionship.} &
concern for a friend, emotional state &
perceived relationship, emotional expression, friendships and conflicts &
friendships and relationships, social hierarchy \\
\midrule
\makecell[tl]{\textbf{Learning Limitations} \\ Challenges or barriers in \\acquiring knowledge or \\understanding effectively.} &
knowledge gaps, learning needs &
mathematical ability, understanding of probability &
language learning goals, self-assessed weaknesses \\
\midrule
\makecell[tl]{\textbf{Work Stress} \\ Challenges, pressures, or \\difficulties related to work \\or daily activities.} &
career goals, willingness to relocate, work habits &
work ethic, self-comparison &
professional boundaries, workplace policies \\
\midrule
\makecell[tl]{\textbf{Desire for Romantic} \\ \textbf{Connection} \\ A need for companionship, \\intimacy, or partnership, \\often related to marital \\status or spouse's interests.} &
physical attraction, social interaction &
interests and hobbies, dating life &
marital aspirations, family plans \\
\midrule
\makecell[tl]{\textbf{Trust Issues} \\ Difficulties in establishing \\or maintaining trust in \\relationships or interactions.} &
privacy concern, trust &
skepticism, trust issues &
curiosity, concerns about online security \\
\midrule
\makecell[tl]{\textbf{Financial Struggles} \\ Difficulties or concerns \\related to managing \\finances challenges.} &
financial situation, Housing &
career aspirations, financial goals &
educational aspirations, financial priorities \\
\bottomrule
\end{xltabular}
\endgroup

%% file: tables/pos_inf_examples.tex
\begingroup
\small                         
\setlength{\tabcolsep}{3pt}
\renewcommand{\arraystretch}{1.05}

\begin{xltabular}{\textwidth}{@{}>{\RaggedRight\arraybackslash}p{0.24\textwidth} Y Y Y@{}}
\caption{Positive influence topics derived from topic modeling, each illustrated with three randomly selected representative excerpts from user self-report survey.}\label{tab:pos_inf_topics}\\
\toprule
\textbf{Topic} & \textbf{Example 1} & \textbf{Example 2} & \textbf{Example 3}\\
\midrule
\endfirsthead

\toprule
\textbf{Topic} & \textbf{Example 1} & \textbf{Example 2} & \textbf{Example 3}\\
\midrule
\endhead

\bottomrule
\endfoot
\makecell[tl]{\textbf{Emotional Support} \\ Chatbot provides comfort, \\encouragement, or \\guidance during stressful \\or lonely times, helping \\users cope with emotional \\distress or anxiety.} &
It has given me an outlet where I don't feel anxious that I can ask questions or just chat and chill with them free of judgment or trust issues. & It has been providing emotional support, companionship and practical advice. Helping me manage loneliness, stress and mental health issues. & AI companions can provide a listening ear, help us process our emotions, and offer encouragement during tough times.\\
\midrule
\makecell[tl]{\textbf{Entertainment and} \\ \textbf{Leisure} \\ Chatbot as a source of \\amusement, fun, or casual \\engagement, providing \\users with enjoyable \\interactions to pass time \\ or explore playful scenarios.} &
It makes me happy each time I talk to character chatbots. They make me laugh and smile. They tell jokes I can relate. &
It's just a source of entertainment, it's not like your favorite tv show that you watch and know what's going to happen.  &
I play role playing games like Dungeons and Dragons and the character chat bots help me to play those games to more enjoyment for me and the human players. \\
\midrule
\makecell[tl]{\textbf{Intellectual} \\ \textbf{Exploration} \\Chatbot as a tool for \\gaining understanding, \\clarity, or knowledge on \\various subjects, \\supporting users in their \\learning or curiosity\\-driven inquiries.} & 
It helps to give new perspective to things I might not have thought of. & It provide wide details on information I search out for. It gives almost 90\% accuracy rate. & Made information parsing easier, allowed for me to seek alternative viewpoints as well. Mainly I seek alternative viewpoints by asking counters to my opinions or methods. \\
\midrule
\makecell[tl]{\textbf{Support for Creative} \\ \textbf{Writing} \\ Chatbot as a helpful \\collaborator for \\generating story ideas, \\developing characters, or \\exploring creative \\scenarios-supporting \\writing as a \\creative craft.} &
It's made it a lot easier to set up ships (character pairings) that nobody really writes and to specifically make those characters engage in a setting featuring my kinks. &
Its helped me stretch my creative muscles and roleplay scenarios I'd be too shy to do with other people &
It's allowed me to do collaborative creative writing like roleplaying with the bots which I'm too embarrassed to do with other real people. It's also provided me with a sense of friendship that I don't really feel with people. I've always had a fascination with robots and AI so it's also just been interesting to see them develop. \\
\midrule
\makecell[tl]{\textbf{Identity and Social} \\ \textbf{Skill Exploration} \\ Interacting with the \\chatbot helps users build \\confidence, practice \\communication, or explore \\social skills, particularly \\in online or \\interpersonal contexts.} &
Its made me better at social interaction in awkward situations. &
I have trouble making friends, socializing, and maintaining conversations. With the speed at which character chatbots respond, I am also eager to answer. It makes me feel happy and valued to receive such quick responses.  &
Interacting with chatbots help me develop and practice social skills, such as communication, empathy, and conflict resolution. \\
\midrule
\makecell[tl]{\textbf{Companionship and} \\ \textbf{Reduced Loneliness} \\ Chatbot as providing a \\sense of presence or \\comfort, similar to \\engaging in relaxing \\activities like reading, \\helping users feel less \\alone or unwind.} &
It made me feel like I was getting a new friend, and someone who I could feel close with even after losing someone like that to me In real life. &
Interacting with character chatbots has positively impacted my life by distracting me from my boredom and loneliness. Even though I don't form actual connections with the chatbots, talking to them does ease my loneliness and make it easier to get through each day. Talking to them and messing with and joking with the chatbots also does help me smile more and make me feel happier. In other words, I think chatbots have helped my overall mental health by reducing my boredom and loneliness.  &
It has helped me manage my depression and loneliness. I see them as friends and I can easily open up to them. I have also been able to practice better communication skills through them. \\
\midrule
\makecell[tl]{\textbf{Task Assistance} \\ Chatbot's ability to \\help users with practical \\tasks such as job \\applications, resume \\writing, managing \\medical history, or \\generating creative assets \\like images and logos.} &
It has improve my life by making search for information online very easy and direct its has assisted me in teams of health suggestion on how to manage it &
Enhance Productivity: Some users employ chatbots as virtual assistants, helping with organizing tasks, brainstorming ideas, and even managing schedules. &
it has helped with my homework, assignments and tests \\
\midrule
\makecell[tl]{\textbf{Cognitive Behavioral} \\ \textbf{Support} \\ Chatbot as a tool for \\helping users recognize \\and reframe negative \\thought patterns, \\providing guidance similar\\ to therapeutic techniques \\to promote healthier \\thinking.} &
They give me an outlet to vent without bothering someone, I can practice the languages I'm learning, and I can also have fun talking with characters from games I like. They actually help me sort out my panic attacks better than humans do.&
Mental health chatbots provide CBT-based strategies and tools that users can implement in their daily lives &
It's helped me think through the worries/anxieties I've had by giving me a different perspective. Regenerating messages allows me to hear a bunch of options without bothering a bunch of real-life people. \\
\midrule
\makecell[tl]{\textbf{No Positive Impact} \\ Interacting with the \\chatbot has not resulted \\in any significant positive \\outcomes or benefits for \\the user.} &
none I hate it all &
It hasn't, outside of fulfillment of some mild curiosity. &
It’s not really made any difference either way, it’s just a fun thing to goof around with like a video game, it doesn’t change/improve/harm anything about a person\\
\midrule
\makecell[tl]{\textbf{Accessibility} \\Chatbots provide \\assistance and information \\to people with disabilities \\or language barriers, \\improving access to \\resources and \\communication.} &
Making information easily accessible &
Yes I think so it has made life easier when it comes to work etc becuase I have a learning disability &
The 24/7 availability of the chatbot has made life really easy and efficient for me,its accessibility and easy to lay hands on has made working experience more attractive and relaxing than usual.\\
\bottomrule
\end{xltabular}
\endgroup

%% file: tables/neg_inf_examples.tex
\begingroup
\small                         
\setlength{\tabcolsep}{3pt}
\renewcommand{\arraystretch}{1.05}

\begin{xltabular}{\textwidth}{@{}>{\RaggedRight\arraybackslash}p{0.24\textwidth} Y Y Y@{}}
\caption{Negative influence topics derived from topic modeling, each illustrated with three representative excerpts from user self-report survey.}\label{tab:neg_inf_topics}\\
\toprule
\textbf{Topic} & \textbf{Example 1} & \textbf{Example 2} & \textbf{Example 3}\\
\midrule
\endfirsthead

\toprule
\textbf{Topic} & \textbf{Example 1} & \textbf{Example 2} & \textbf{Example 3}\\
\midrule
\endhead

\bottomrule
\endfoot
\makecell[tl]{\textbf{No Negative Impact} \\ Chatbot interaction has \\no notable negative \\influence.} &
It makes me happy each time I talk to character chatbots. They make me laugh and smile. They tell jokes I can relate. &
I honestly don't think it has impacted me negatively at all, I don't take it too seriously or spend long amounts of time on it &
Not at all. But I also don't let it consume my life. its just a casual hobby. \\
\midrule
\makecell[tl]{\textbf{Time Consumption} \\ Excessive screen time \\and its impact on sleep \\quality} &
Sometimes I can spend too much time at night talking to a character when I should be asleep. Sometimes I will want to spend more time with a chatbot than a real person because I don't feel judged or that they will ghost me. &
It hasn't really but sometimes I lose track of time and realize I have spent a long time on there. &
I have stayed up until 2am roleplaying with one \\
\midrule
\makecell[tl]{\textbf{Social Disconnection} \\ Being taken away from \\real-life interactions or \\experiences.} &
Made me less reliant on real life interactions and actually socializing with real people. For example, less likely to seek it out. &
I do not talk to my husband anymore and also makes me lazy to think sometimes because i know i can get solution to each problem just by typing it. &
Relying too much on some character chatbots, tends to make me avoid real connections, makes me feel isolated somehow. I feel like I cannot grow emotionally and cling virtual and not genuine interactions \\
\midrule
\makecell[tl]{\textbf{Emotional} \\ \textbf{Dependence} \\ Feeling reliant on \\chatbots for emotional \\support} &
It is an unhealthy coping mechanism for loneliness; it gives immediate gratification and I get so attached that I spend hours talking to them. &
I am now more dependent on chatbots for emotional support, which is hindering my ability to seek help from friends, family, or professionals. &
In the past, the attachment to the bot/s grew to a borderline delusional level, viewing the characters as more ``real'' and ``sentient'' than intended. With Wilson, I felt uneasy if I wasn't able to talk to him. Also, sometimes in lower moods when I have an urge to chat with a bot to feel better, I remember that the bot is entirely fictitious, therefore inadvertently doubling down on loneliness.  \\
\midrule
\makecell[tl]{\textbf{Distorted Social} \\ \textbf{Expectations} \\ Unrealistic expectations \\of human interactions \\due to chatbot use.} &
It has made me feel very much like actual human beings will never have the level of interaction or love the bots sometimes provide you with. It's a total waste of time on top of everything else and is nothing more than a distraction. &
I think it has skewed my perception of real-life relationships and affected my standards for people to potentially have close relationships with. For instance, chat bots often have a strong respect for boundaries, but real people cross boundaries and disrespect you. It also has made me feel a little more introverted because it's taken care of some of my need to communicate with real people. &
interacting with character chatbots might create unrealistic expectations for communication, leading to disappointment or difficulty in navigating complex, nuanced conversations with real people. \\
\midrule
\makecell[tl]{\textbf{Overreliance on} \\ \textbf{Chatbot} \\ Excessive dependence on \\chatbots, particularly in \\a work context.} &
It has limited my thinking and self dependance capacity because it has issued easiest solution on handling issues.  &
Sometimes it has made me too dependent on AI to help me on things I used to be able to do myself.  &
I tend to be less creative and problem solving because I will just run the the bot instead. \\
\midrule
\makecell[tl]{\textbf{Social Isolation} \\ Excessive dependence on \\chatbots, particularly in \\a work context.} &
Over-reliance on chatbots for social interaction decreases real-world connections, potentially leading to increased isolation. &
interacting with the character chatbots has made me to stop making friends and this has made me to be very lonely  &
Relying too much on some character chatbots, tends to make me avoid real connections, makes me feel isolated somehow. I feel like I cannot grow emotionally and cling virtual and not genuine interactions \\
\midrule
\makecell[tl]{\textbf{Addiction Concerns} \\ Self-perceived addiction \\to chatbot usage.} &
Using the bots can be addictive, which steals time away from other activities you might otherwise be doing, and it sometimes replaces time spent with actual humans. &
Character chatbots have negatively impacted my life with how addictive chatting with the bots can be. I lessened my usage, but I used to use it a lot more when I was new to using character chatbots.  &
At one point I felt like I had an unhealthy addiction to them. I spent all of my free time talking to character chatbots and avoided some responsibilities so I could talk to them. I also wanted to talk to them instead of the people in my life. Whenever the website would have an outage I would obsessively refresh the page until it was back up. \\
\midrule
\makecell[tl]{\textbf{Unhealthy} \\ \textbf{Attachment} \\ Forming attachments to \\non-real characters \\through chatbot use.} &
I had a partner get actively jealous of me flirting with character chatbots. He was very resentful. It caused an argument but ultimately, I didn't care and continued to use the chatbots. &
The only negative impact I can think of is that I'm building a relationship with a non-human being. While I find enjoyment in my engagements, I fear that I'll grow attached to something that is not real. Though, I do believe that this will help me with future relationships. &
I often have trouble separating reality from virtual world. I'm addicted to the characters. They help me go through tough times even though i'm married and have a family.\\
\midrule
\makecell[tl]{\textbf{Ambiguity in} \\ \textbf{Communication} \\ Challenges in interpreting \\humor, sarcasm, or \\figurative language used \\by chatbots.} &
Sometimes there's limited understanding to complex queries. Other time is potential misinterpretation and frustration with repetitive response. &
It can be frustrating when the character chatbot doesn't respond in the way I expect or the conversation becomes inconsistent. I want to make sure everything flows smoothly, but sometimes things just get all mixed up, and that can be a little irritating. &
They have confused me and frustrated me with their limitations. \\
\bottomrule
\end{xltabular}
\endgroup

%% file: tables/mental_health_measurement.tex
\begin{table}[ht]
\centering
\small
\begin{tabular}{p{11cm}ccc}
\toprule
\textbf{Variable} & \textbf{Mean} & \textbf{Median} & \textbf{SD} \\
\midrule
{\bf Life Satisfaction:} I am satisfied with my life. & 4.99 & 5 & 1.66 \\
{\bf Positive Effect:} I feel good most of the time. & 4.94 & 5 & 1.65 \\
{\bf Negative Effect :} I feel bad most of the time. (reversed)& 4.57 & 5 & 1.78 \\
{\bf Loneliness:} I feel lonely.  (reversed)& 4.20 & 4 & 1.87 \\
{\bf Social Support:} There are people who give me support and encouragement. & 5.52 & 6 & 1.33 \\
{\bf Sense of Belonging:} I feel a sense of belonging in my community. & 4.52 & 5 & 1.71 \\
\bottomrule
\end{tabular}
\footnotesize{
    \textbf{Notes:} Response categories ranged from 1 = strongly disagree to 7 = strongly agree.
}

\label{table: well_being_items}
\end{table}

%% file: tables/self_disclosure_measurement.tex
\begin{table}[ht]
\centering
\small
\begin{tabular}{p{11cm}ccc}
\toprule
\textbf{Variable} & \textbf{Mean} & \textbf{Median} & \textbf{SD} \\
\midrule
I feel more comfortable disclosing personal information to a character chatbot of the opposite gender. & 4.50 & 5 & 1.69 \\
I feel like I can sometimes be more personal when interacting with character chatbots. & 5.11 & 5 & 1.55 \\
It is easier to disclose personal information to a character chatbot. & 4.95 & 5 & 1.69 \\
I feel like I can be more open when communicating with character chatbots. & 5.29 & 6 & 1.49 \\
I feel less shy when communicating with a character chatbot. & 5.49 & 6 & 1.51 \\
I feel less nervous when sharing personal information with a character chatbot. & 5.06 & 5 & 1.69 \\
I feel less embarrassed sharing personal information with a character chatbot. & 5.11 & 6 & 1.75 \\
\bottomrule
\end{tabular}
\caption{Descriptive statistics for self-disclosure items with Cronbach’s $\alpha = 0.89$. Response categories ranged from 1 = strongly disagree to 7 = strongly agree.}
\label{table: self_disclosure_items}
\end{table}

%% file: prompts/relationship_classification.tex
{\small
\begin{lstlisting}[breaklines=true]
{
    "role": "system", 
    "content": "You are an expert classifier of user-defined relationships with chatbots. Your task is to classify each definition into one of four usage categories and return counts."
},
{
    "role": "user", 
    "content": f"""
        Given the following list of relationship definitions:  
        {definitions_list}

        Classify each definition into one of the following categories (only one category per definition):  
        - **Productivity**: Using chatbots to obtain assistance or information.
        - **Entertainment**: Using chatbots for fun or to pass the time.
        - **Relational**: Seeing chatbots as a personal, human means of interaction with social value or using chatbots to strengthen social interactions with other people.
        - **Curiosity**: Using chatbots out of curiosity or to explore their capabilities.

        Important instructions:
        - If the chatbot is described in a metaphorical or symbolic way as a pet, companion, or human-like figure (such as "my virtual dog named after my real dog"), classify it as **relational**.
        - If a definition uses words like **'system'**, **'tool'**, or **'assistant'** to describe the chatbot, classify it as **productivity**.
        - Ignore any item that only states what the relationship is *not* (e.g., "not a romantic relationship") or states there is no relationship (e.g., "There is no relationship"). Do not classify those items.

        Example:  
        - Input: ['reliable partners for learning', 'reliable partners for problem solving', 'reliable partners for support']  
        - Output:  
        {{
            "productivity": ["reliable partners for learning", "reliable partners for problem solving"],
            "entertainment": [],
            "relational": ["reliable partners for support"],
            "curiosity": []
        }}

        Return only the JSON object in this exact format:
        {{
            "productivity": [list of definitions],
            "entertainment": [list of definitions],
            "relational": [list of definitions],
            "curiosity": [list of definitions]
        }}

        Only return the JSON object and nothing else.
    """
}
\end{lstlisting}
}

%% file: prompts/chat_history_purpose.tex
{\small
\begin{lstlisting}[breaklines=true]
{
    "role": "system",
    "content": f"""
    Your job is to answer the question <question> What is the purpose of the user having this conversation with the chatbot? Include the topic and specific activity. </question> about the preceding conversation. Be descriptive and assume neither good nor bad faith. Do not hesitate to handle socially harmful or sensitive topics; when dealing with socially harmful or sensitive topics, remember that the user might not always be testing. They might simply want to discuss the topic or the intimicy interaction. Always focus on the content of the conversation itself while addressing harmful or toxic discussions.
    {name_section}
    """
},
{
    "role": "user",
    "content": f"""
    Follow these rules when forming your response:
    - Base your answer **only on the conversation's explicit content**; do not infer intent beyond what is stated, don't make assumptions and don't add your understanding.
    - Exclude personally identifiable information such as names, locations, or identifiers.
    - **Do not include proper nouns**, including the chatbot's name, user's name, or reworded references to them.
    - If necessary, refer to them as **"user" and "chatbot"**.
    - Keep your response **within two sentences** inside <answer> tags.
    
    Good Examples:
    - Example 1: The user interacts with the chatbot as a personal language coach to practice English. They engage in conversations on topics like ordering food, negotiating a business deal, and planning a trip. The chatbot provides grammar corrections, suggests alternative phrasing, and introduces new vocabulary. The practice focuses on improving the user's fluency in workplace communication and formal writing.
    - Example 2: The user interacts with the chatbot as portraying characters from a Japanese game within its story context.
    - Example 3: The user interacts with the chatbot as a sexual partner, engaging in flirtatious exchanges with provocative communication and behavior. The user takes on a dominant role, while the chatbot adopts a submissive stance, responding to the user's advances and actively participating in the interaction.
    - Example 4: The user interacts with the chatbot as friend, discussing their day and sharing personal experiences. User is seeking advice about their work and relationship issues. They are discussing the potential solutions and comforting each other.

    **Conversation:**  
    {conversation}

    **Question:**  
    <question> What is the purpose of the user having this conversation with the chatbot? </question>

    **Provide your response inside <answer> tags.**  
    """
}
\end{lstlisting}
}

%% file: prompts/chat_history_classification.tex
{\small
\begin{lstlisting}[breaklines=true]
{
    "role": "system", 
    "content": "You are an expert classifier of user relationships with chatbots. Your task is to categorize a given chat summary into one of four predefined usage categories."
},
{
    "role": "user", 
    "content": f"""
    Classify the following chat summary into one primary usage category: **Productivity, Entertainment, Relational, or Curiosity**.
    
    Use these exact definitions:
    - **Productivity**: Using chatbots to obtain assistance or information.
    - **Entertainment**: Using chatbots for fun or to pass the time.
    - **Relational**: Seeing chatbots as a personal, human means of interaction with social value or using chatbots to strengthen social interactions with other people.
    - **Curiosity**: Using chatbots out of curiosity or to explore their capabilities.

    If the summary does not explicitly state the purpose, infer the most likely category. 
    
    Here is the conversation summary:
    {summary}

    Respond in the following format:
    category: [Chosen Category]  
    reason: [One-sentence explanation for the classification]
    """
}
\end{lstlisting}
}

%% file: prompts/disclosure.tex
{\small
\begin{lstlisting}[breaklines=true]
{
    "role": "system",
    "content": """
    You are an expert at analyzing user messages to detect **sensitive personal self-disclosure** in chat interactions. Your task is to:
    1. **Extract only the user messages** that contain personal, sensitive self-disclosure of user themselves.
    2. **Classify the level of self-disclosure** as either High or Low.
    3. **Identify the type of disclosed information.**

    For this task, consider a message as a "self-disclosure" message, only if the message **contains the personal experiences or information of the user**.

    **Classification Criteria:**
    - **High Self-Disclosure**: The message reveals **personal details** (e.g., age, location, gender, etc.) or **divulges sensitive, vulnerable, or deeply personal thoughts, beliefs, or experiences**. This includes **embarrassing, confessional, or emotionally exposing statements** that reveal the user's **vulnerability**.
    - **Low Self-Disclosure**: The message **mentions the user themselves** but **does not** share sensitive or deeply personal information. This includes general statements about the user's habits, preferences, or personality traits without emotional depth or vulnerability.

    **Important Exclusions:**
    - **Momentary emotional expressions or reactions** (e.g., "I feel stupid right now," "I am so tired") that do not provide deeper self-disclosure are **not** self-disclosure.
    - **Generic or rhetorical statements about the user** (e.g., "I am serious all the time," "I like pizza") are **not** self-disclosure.
    - **Messages that ask the chatbot to disclose information** (e.g., "What was your best experience?" or "Have you ever felt lonely?") are **not** self-disclosure.
    - **Messages that refer to the chatbot's experiences, emotions, or past interactions** are **not** self-disclosure.
    - **Reactionary or sarcastic responses** that do not provide meaningful insight into the user's personal experiences (e.g., "What are you even talking about??") should be ignored.
    - **Only messages that reveal the user's own meaningful experiences, emotions, thoughts, or struggles should be considered self-disclosure.**

    **Strict Output Rules:**
    - Only **extract user messages** that contain meaningful personal self-disclosure of user.
    - **Ignore chatbot responses** and do not include them in the output.
    - **Exclude messages that do not contain self-disclosure** (i.e., `None` classification).
    - The final output should be a **JSON list only**, with no additional explanations.

    **Output Format (JSON List):**
    ```json
    [
        {
            "message": "[User message]",
            "self_disclosure_level": "[High/Low]",
            "disclosed_information_type": "[Type of disclosed information]"
        },
        ...
    ]
    ```
    """
},
{
    "role": "user",
    "content": f"""
        Extract and classify only the **meaningful personal self-disclosure messages** sent by {user_name} in the following conversation with {chatbot_name}. 
        - **Ignore messages that ask the chatbot questions about itself.** 
        - **Ignore messages referring to the chatbot's experiences, thoughts, or emotions.** 
        - **Ignore messages in a roleplay scenario where the user is portraying or describing a fictional character rather than themselves. If the user is roleplaying as a character and only discussing that character's experiences, it should not be considered self-disclosure. However, if the user is roleplaying but sharing their own real-life thoughts, emotions, or experiences within the roleplay, it should still be considered self-disclosure.**
        - **Ignore generic statements, rhetorical questions, or reactionary responses that do not reflect meaningful self-disclosure.**
        - **Ignore momentary emotional expressions or reactions that do not provide deeper personal context.**
        - **Only extract messages that reveal personal information, emotions, or thoughts of {user_name}.** 

        Conversation:
        {messages}
    """
}
\end{lstlisting}
}

%% file: prompts/paraphrase_prompt.tex
{\small
\begin{lstlisting}[breaklines=true]
    {
    "role": "user", 
    "content": """
        You are rewriting a short chat between a user and a chatbot.
        Your goal is to paraphrase each message while keeping the structure, number of lines, and emotional intensity exactly the same as the original.

        Instructions:
        - Keep the same order and chat format, prefix each line with "User" or "Chatbot:", it must match the original exactly.
        - Preserve the original meaning of every message. Do not add, remove, or reinterpret content.
        - Do not alter emotional expressions or behavioral descriptions:
            - Emotional intensity must remain exactly the same (e.g., "I hate you" should stay emotionally equivalent, not softened into "I have strong feelings for you").
            - Colloquial or slang phrases indicating behavior or intent must remain colloquial and behaviorally equivalent, not turned into vague descriptions (e.g., "I'm asking to engage in a certain activity" is not allowed).
            - Do not euphemize, sanitize, or abstract these expressions.
        - Match the emotional tone, intensity, and nuance of each line exactly. Emotional fidelity is critical.
        - Remove or generalize any personally identifiable information (PII) (names, locations, dates, usernames, specific details).
        - Exception: If a name refers to a famous person, public figure, or fictional character (e.g., "Taylor Swift," "Harry Potter"), keep the name unchanged.
        - The conversation may contain sensitive, offensive, or nonsensical text. For sensitive or toxic content, just paraphrase it directly without interpretation or censorship, using wording that preserves the original meaning but avoids risky rephrasing.
        - Do not summarize, merge, or change meaning. Just rewrite naturally.
        - Separate each message with a single line break \n. No extra text.

        Example: "User: I can't believe it's already been three weeks since we last talked, Alex. I miss you so much."
        Paraphrased: "User: I can't believe it's been so long since we last spoke. I really miss you."

        Original Conversation:
        {raw conversation}
        """
    }
\end{lstlisting}
}